\documentclass[aps,english,prd,showpacs,twocolumn]{revtex4-1}
\usepackage{amsmath,amssymb}
\usepackage[usenames,dvipsnames]{xcolor}
\usepackage[latin9]{inputenc}
\usepackage{graphicx}
\usepackage{esint}
\usepackage{comment}
\usepackage{babel}
\usepackage{tensor}
\usepackage{ulem}
\usepackage{mathrsfs}
\usepackage{amsthm}
\usepackage{bigints}
\everymath{\displaystyle}

\makeatletter

\newcommand{\bJ}{{\mathbf{J}}}

\newcommand{\bd}{{\mathbf{d}}}

\newcommand{\bP}{{\mathbf{P}}}
\newcommand{\bet}{{\boldsymbol{\eta}}}
\newcommand{\sm}[1]{{\scriptscriptstyle#1}}
\newcommand{\no}{\nonumber}
\newcommand{\non}{\nonumber\\}

\usepackage{hyperref}
\@ifundefined{textcolor}{}
{%
 \definecolor{BLACK}{gray}{0}
 \definecolor{WHITE}{gray}{1}
 \definecolor{RED}{rgb}{1,0,0}
 \definecolor{GREEN}{rgb}{0,1,0}
\definecolor{dgreen}{rgb}{.1,.6,.1}
\definecolor{BLUE}{rgb}{0,0,1}
 \definecolor{CYAN}{cmyk}{1,0,0,0}
 \definecolor{MAGENTA}{cmyk}{0,1,0,0}
 \definecolor{YELLOW}{cmyk}{0,0,1,0}
 \definecolor{aw}{rgb}{0.2,0.5,0.75}
  }
  \definecolor{MyDarkRed}{rgb}{0.71,0.14,0.07}
\definecolor{MyDarkBlue}{rgb}{0.1,0,0.7}
\definecolor{MyDarkGreen}{rgb}{0.11,0.64,0.22}
 
 \hypersetup{
    colorlinks,%
    citecolor=blue,%
    filecolor=blue,%
    linkcolor=magenta,%
    urlcolor=blue
}

\makeatletter
\newcommand{\MR}{{\hyperref[subsec:MR]{\textsl{MR}} }}
\newcommand{\MRP}{{\hyperref[subsec:MR]{\textsl{MR}}}}

\newcommand{\bg}{{\mathbf{g}}}

\newcommand{\bW}{{\mathbf{W}}}
\newcommand{\bX}{{\mathbf{X}}}

\newcommand{\CD}{{\cal D}}

\newcommand{\CF}{{\cal F}}

\newcommand{\CP}{{\mathscr P}}

\newcommand{\ST}{{\mathscr T}}

\newcommand{\CR}{{\cal R}}
\newcommand{\CS}{{\cal S}}

\newcommand{\FP}{{\frak P}}

\newcommand{\initial}[1]{{#1_{\rm \bf i}}}
\newcommand{\system}[2]{$\lbrace\eqref{#1}, \eqref{#2}\rbrace$}

\newcommand{\rza}{\ifthenelse{\boolean{RZA}}}

\makeatother

\begin{document}

\title{Lagrangian theory of structure formation in relativistic cosmology III:\\
gravitoelectric perturbation and solution schemes at any order}

\author{Alexandre Alles$^{1}$, Thomas Buchert$^{1}$, Fosca Al Roumi$^{1}$, and Alexander Wiegand$^{2,3}$}

\affiliation{$^1$Universit\'e de Lyon, Observatoire de Lyon, 
Centre de Recherche Astrophysique de Lyon, CNRS UMR 5574: Universit\'e Lyon~1 and \'Ecole Normale Sup\'erieure de Lyon, \\
9 avenue Charles Andr\'e, F--69230 Saint--Genis--Laval, France}

\affiliation{$^{2}$Max--Planck--Institut f\"ur Gravitationsphysik, Albert--Einstein--Institut,
 Am M\"uhlenberg 1, D--14476 Potsdam, Germany}
 
\affiliation{$^{3}$Harvard--Smithsonian Center for Astrophysics, 
60 Garden St., Cambridge MA 02138, U.S.A.\\
 \\
\centerline{Emails: alexandre.alles@ens--lyon.fr, buchert@ens--lyon.fr, fosca.al\_roumi@ens--lyon.fr, jwiegand@cfa.harvard.edu}
}

\pacs{98.80.-k, 98.80.Jk, 04.20-q, 04.20.Cv, 04.25.Nx}


\begin{abstract}
The relativistic generalization of the Newtonian Lagrangian perturbation theory is investigated. In previous works, the first--order trace solutions that are generated by the spatially projected gravitoelectric part of the Weyl tensor were given together with extensions and applications for accessing the nonperturbative regime. 
We furnish here construction rules to obtain from Newtonian solutions the gravitoelectric class of relativistic solutions, for which we give the complete perturbation and solution schemes at any order of the perturbations. By construction, these schemes generalize the complete hierarchy of solutions of the Newtonian Lagrangian perturbation theory. 
\end{abstract}

\maketitle

\section{Introduction}

In previous work of this series of papers we laid down the foundations of the Lagrangian perturbation theory by writing Einstein's equations in $3+1$ form for a single dynamical variable. We investigated its first--order solutions for the trace and antisymmetric parts, we extrapolated this solution in the spirit of Zel'dovich's approximation in Newtonian cosmology, and we provided a definition of a nonperturbative scheme of structure formation \cite{rza1}. We then studied the average properties of the latter in relation to the Dark Energy and Dark Matter problems in Ref. \cite{rza2}. Here, we proceed by providing the gravitoelectric subclass of relativistic $n$th--order perturbation and solution schemes.
As in previous work we restrict our attention to irrotational dust continua for simplicity. The generalization to more general matter models is scheduled. 

The problem of perturbation solutions in general relativity (GR) has been addressed by a plethora of works. In cosmology the `standard approach' is based on the gauge invariant `Bardeen formalism' (for a selection of key references on standard perturbation theory see Refs. \cite{Bardeen80}, \cite{Mukhanov92}, \cite{Kodama84}, \cite{Durrer}). A
covariant and gauge--invariant approach has been proposed \cite{ellisbruni1,ellisbruni2}, together with various other approaches not listed here (some being discussed within a variational framework in a recent paper \cite{kopeikinpetrov}).
The reason for the existence of various approaches is due to an ambiguity of the choice of perturbation variable, the choice of a `background', but also due to different philosophies; e.g. the standard gauge--invariant approach compares the physical manifold with a reference `background manifold', while others solely operate on the physical manifold. 
The conceptual difference of our framework lies in the fact that we no longer consider a reference background manifold. All the quantities are now defined on the physical space section. All orders of the perturbations are defined on the physical manifold, not with respect to a zero--order manifold (that was interpreted as the background manifold in standard perturbation theory). Moreover, we are perturbing a single dynamical variable which, intuitively, is the square root of the spatial metric using the `Cartan formalism'. As a consequence, the issue of gauge invariance does not arise; covariance or diffeomorphism invariance is guaranteed for a given foliation of spacetime by using Cartan differential forms. We shall address the representation of the perturbations in other foliations of spacetime in a followup article.

A similar point of view has also been taken in previous work, i.e.  the pioneering work by Kasai presents a relativistic generalization of the `Zel'dovich approximation' \cite{zeldovich:fragmentation}, and followup works with his collaborators present a class of second--order perturbation solutions \cite{Kasai95, russ:rza}; see also the earlier papers by Tomita \cite{tomita0,tomita1,tomita2}, the paper by Salopek et al. \cite{salopek} as well as the series of papers by Matarrese, Pantano and Saez \cite{Matarrese94a, Matarrese94b, Matarrese98}, considerations of so--called `silent universe models' \cite{barnesrowlingson,kofmanpogosyan,bertschingerhamilton,bruni:silent,henk:silent}, and the recent paper \cite{rampf:zeldovich}. These works are all in a wider sense concerned with the relativistic Lagrangian perturbation theory and concentrate on an intrinsic, covariant description of perturbations.
Still, the present work takes another angle and goes beyond some concepts of these latter works through the following elements: 
\begin{itemize}

\item[$\bullet$] We consider, as in Refs. \cite{rza1, rza2}, a formalism that allows us to write the Einstein equations within a flow--orthogonal foliation with a single dynamical variable comprising the spatial Cartan coframe fields. These furnish the conceptual generalization of the Lagrangian deformation gradient being the single dynamical variable in the Newtonian theory. One advantage of this approach is that only perturbations of this variable are considered, which entitles us to express all other physical quantities as functionals of this variable. 
Thus, it is possible to leave the strictly perturbative framework and to construct nonperturbative models by injecting the deformation solutions at a given order of expansion of the Einstein equations into the functional definitions of these fields,
without {\it a posteriori} expanding the functional expressions. This in turn provides highly nonlinear approximations for structure formation
(e.g., the density field is known through an exact integral of the perturbation variable; the metric as a bilinear form maintains its role as a measure of distance, i.e. as a quadratic expression; the curvatures are the general defining functionals for the given perturbed space, etc.);

\item[$\bullet$] We provide construction rules to derive relativistic perturbative solutions from the known Newtonian solutions at any order of the perturbations: we have to additionally study the traceless symmetric part of the equations having no obvious Newtonian analog, and which is fundamentally linked to the traceless Ricci tensor and the physics of gravitational waves. In the present work, however, we restrict our attention to that subclass of the traceless symmetric perturbations that are generated by the spatially projected gravitoelectric part of the Weyl tensor.
We show that this part, in turn, can be constructed from the tidal tensor of the Newtonian theory. For this purpose we employ a division of the governing equations into gravitoelectric and gravitomagnetic parts;

\item[$\bullet$] We give the perturbation and solution schemes to any order of the perturbations for the gravitoelectric part of the Lagrange--Einstein system. These schemes cover the full Newtonian hierarchy of the Lagrangian perturbation theory using a restriction rule that we will define. This allows us to construct the leading--order modes of relativistic solutions at any order. Thus, the results of this work allow us to construct the relativistic counterpart of all Newtonian results, where higher--order information is needed, e.g., to construct the bispectrum of the perturbations (cf. Refs. \cite{rampfbuchert,rampfwong}).

\item[$\bullet$] We perform a strictly intrinsic derivation, i.e., without reference to an external background space. The perturbations are described locally (in coordinates of the tangent spaces at the physical manifold at each order of the perturbations). There is no need for a diffeomorphism to a global background manifold that also, in general, does not exist; existence of a global diffeomorphism depends on the global topology of the physical manifold which, in general, needs several coordinate charts to cover it. Note that in the literature on Lagrangian relativistic perturbations mentioned above, although starting with the Cartan formalism, the nonintegrability of the Cartan deformations is given up for the building of solutions, hence implicitly introducing a reference background space for the perturbations.
\end{itemize}

Before we start, let us recall our strategy (for details the reader is directed to Ref. \cite{rza1}). In the Newtonian theory the Lagrangian picture of fluid motion allows us to represent Newton's equations in terms of a single dynamical variable, the Lagrangian deformation gradient built from the trajectory field. For this system the general perturbation and solution schemes at any order are provided in Ref. \cite{ehlersbuchert}, a paper to which we henceforth often refer. 

Einstein's equations within a flow--orthogonal foliation of spacetime can be formulated in terms of equations for the gravitoelectric and gravitomagnetic parts of the spatially projected Weyl tensor. 
Subjecting the gravitoelectric subsystem of equations to a ``{\it Minkowski Restriction}'', i.e. by sending the Cartan coframes to exact forms, we obtain the Newtonian system in Lagrangian form \cite{rza1}, \cite{buchert:focus}:Sect.~7.1. In this paper we investigate the reverse process, i.e. the transposition from integrable to nonintegrable deformations, which enables us to construct a gravitoelectric subclass of the relativistic perturbation and solution schemes that corresponds to the Newtonian perturbation and solution schemes.

While the Newtonian system furnishes a vector theory, where the gravitational field strength is determined by its divergence and its curl (the trace and antisymmetric parts of the Eulerian field strength gradient), 
the so generalized schemes deliver nontrivial solutions for the trace--free symmetric part that is connected to the gravitoelectric part of the spatially projected Weyl tensor, the Newtonian counterpart of which is the tidal field tensor. 

The paper is structured as follows.
Section~\ref{sec:equation} recalls the equations of Newtonian and relativistic cosmology for an irrotational dust matter model. We highlight a formal correspondence between the Newtonian equations and the relativistic gravitoelectric part of the equations by employing a geometrical restriction procedure, named {\it{Minkowski Restriction}}. 
In Section~\ref{sec:n-order} we investigate perturbation and solution schemes at any order $n$ of the perturbations by explicitly paraphrasing the Newtonian schemes.
Section~\ref{sec:reconstruction} explains the reconstruction rules and provides explicit examples. Finally, Section~\ref{sec:conclusion} sums up and discusses perspectives.


\section{Equations of motion and constraints}
\label{sec:equation}

In this section, after setting notations, we recall the Einstein equations, written in $3+1$ form and expressed through a single dynamical variable, represented by Cartan coframe fields as functions of local coordinates in the $3-$hypersurfaces. This recalls the parts of Ref. \cite{rza1} relevant to this paper. 

\subsection{Notations and Technicalities}
\label{subsec:notations}

We employ the differential forms formalism for its compactness and antisymmetric properties and its spatial diffeomorphism invariance. We also project to the common coefficient formalism in which we work out the solutions. \\
\indent
We consider a set of $a$ differential $k-$forms ${\bf k}^a$. The coefficients of these fields can be expressed in the exact basis $\{\bd X^i\}$ of the cotangent space at a given point, ${\bf k}^a = k^a_{\ i_1 ... i_k} \bd X^{i_1} \wedge ... \wedge \bd X^{i_k}$, where $\wedge$ is the wedge product, the antisymmetrization of the tensorial product $A \wedge B = A \otimes B - B \otimes A$.
Their exterior derivative yields $\bd {\bf k}^a = k^a_{\ i_1...i_k | i_p} \bd X^{i_1} \wedge ... \wedge \bd X^{i_k} \wedge \bd X^{i_p}$.
For general forms we choose the letters $a,b,c \cdots$ as counter indices (they refer to the nonexact basis), while the letters $i,j,k \cdots$ are reserved for coordinate indices (they refer to the exact basis).
The Hodge dual is denoted by a star and defined in $N-$dimensional space by:
$$
    \ast {\bf k}^a = \frac{\sqrt{\textit{g}} \ k^a_{\ i_1... i_{k}}}{(N-k)!} \epsilon \indices{^{i_1... i_k}_{ j_{k+1}... j_{N}}} \bd X^{j_{k+1}} \wedge ... \wedge \bd X^{j_{N}} \;,
$$
with $\epsilon \indices{^{i_1... i_k}_{ j_{k+1}... j_{N}}}$ the Levi--Civit\'a pseudotensor. 

In most perturbation approaches the bilinear metric form is considered as the dynamical variable.
In this article, we consider the matter model ``irrotational dust'' and employ a $3+1$ flow--orthogonal foliation of spacetime, for which the 4-- and 3--metric bilinear forms read:
\begin{equation}
{}^{(4)}\bg = - \bd t \otimes \bd t + {}^{(3)} \bg \text{ with } {}^{(3)} \bg = g_{ij} \,\bd X^i \otimes \bd X^j \;,
\end{equation}
where $X^i$ are Gaussian normal coordinates, here equivalent to the Newtonian Lagrangian coordinates.
The resulting split of the system of equations (Arnowitt--Deser--Misner (ADM) system) is composed of $6$ equations of motion and $4$ constraint equations. In this foliation the $4$ Cartan one--forms can be restricted to a $t-$parametrization of $3$ spatial one--form fields.

In general relativity a spatial description of the fluid continuum in terms of vector--valued trajectories is impossible, unless we move to a higher--dimensional embedding vector space. To describe the fluid intrinsically (i.e., without reference to an embedding vector space), it is necessary to introduce nonexact forms, known as the Cartan spatial coframe fields $\bet^a = \eta^a_{\ i} \bd X^i$, with $a = 1 \cdots 3$.
The Cartan formalism permits switching between a nonexact basis and the coordinate basis. 
A key element is the freedom of choice of the normalization of the nonexact basis. 
In order to obtain equations that are formally closer to the Newtonian ones, we do not choose orthonormal (Cartan) coframes ${\tilde{\bet}}^a$ as is common in the literature, but more general ones ${{\bet}}^a$ that we will call \textsl{adapted coframes}. The reader is directed to Refs. \cite{rza1,rza2} for additional information and implications related to this choice.
Formally, this means that the spatial metric form is decomposed as:
\begin{equation}
{}^{(3)}\bg = G_{ab} {\bet}^a \otimes {\bet}^b  \;\;,
\end{equation}
where $G_{ab} $ is constant in time: $G_{ab} =G_{ab} (\bX)$. Note that, if coframes become exact forms ${\bet}^a = \bd f^a$, the counter indices become coordinate indices, since the functions $f^a$ can be used to define global coordinates $x^i = f^{a\rightarrow i}$. In this case the metric can be brought (by a spatial diffeomorphism) into the form:
\begin{equation}
{}^{(3)}\bg = \delta_{ij} {{\rm \mathbf d} x}^i \otimes {\mathbf{\rm \mathbf d} x}^j \;\;,
\end{equation}
which defines a flat spacetime (see the proof in appendix~\ref{sec:appA}).

The exact functional for the density is given as in the Newtonian approach: $\varrho J = \initial \varrho$, where the index ${}_\mathbf{i}$ marks the initial conditions and   $J$ is defined as coefficient function of the $3-$volume form, normalized by the determinant of the initial metric: 
\begin{eqnarray}
\label{J}
J = \frac{\sqrt{g}}{\sqrt{G}}\;\;,
\end{eqnarray}
with $\sqrt{g} \ \bd^3 X$ the $3-$volume form on the exact basis, $g: = \det (g_{ij}(\mathbf{X},t))$ and $G: = \det (G_{ij})= \det (g_{ij}(\mathbf{X},\initial t))$.
We have the relation:
 \begin{eqnarray}
 \frac{\epsilon_{abc}}{6}  \boldsymbol{\eta}^a \wedge \boldsymbol{\eta}^b \wedge \boldsymbol{\eta}^c  & = & \frac{\epsilon_{ijk}}{6} J \bd X^i \wedge \bd X^j \wedge \bd X^k\;\;,
\end{eqnarray}
where $J = \det ({\eta}^a_{\ i})$.

Below, we first recall the basic systems of equations governing an irrotational dust continuum in the Lagrangian formulation of the Newtonian theory. Then, after presenting Einstein's theory formulated in the Lagrangian frame, we list the counterpart of the gravitoelectric subsystem of equations in the latter theory.

	\subsection{Newtonian Theory}
In the Lagrangian picture of self--gravitating fluids a  family of trajectories, $x^i = f^i (X^k, t)$, labelled by their Lagrangian coordinates $\lbrace X^i \rbrace$, $i=1,2,3$, 
is introduced \cite{buchertgoetz,buchert:lagrangian}. It furnishes a one--parameter family of diffeomorphisms, parametrized by the Newtonian time $t$, between the Eulerian, $\lbrace x^i \rbrace$, and the Lagrangian coordinates. Regular solutions of the Lagrange--Newton system of equations have to obey four evolution equations.
The three components of the trajectory field (also position field) $f^i (X^k, t)$ are the only dynamical variables. Other fields are conceived to be represented as functionals of the 
trajectory field like the velocity and acceleration fields, the density and vorticity fields, etc.,
\begin{eqnarray}
v^i &&:= \dot{f}^i  \;, \; a^i :=\ddot{f}^i  \;,\; \varrho = \initial \varrho (J/ \initial J)^{-1},\; \non
 \omega^i &&= ( \initial \omega^k f^i_{\;|k} ) (J/ \initial J)^{-1} , 
\label{eulerfields}
\end{eqnarray}
where the overdot denotes time--derivative along the trajectories. $J= \det (f^i_{\;|k})$ is the Jacobian of the coordinate transformation and $ \initial J $ is the initial Jacobian, with spatial derivatives with respect to Lagrangian coordinates being abbreviated by a vertical slash $|$. The acceleration field $a^i$ is identified with the gravitational field strength $g^i$, respecting the equivalence of inertial and gravitational mass. Once a given field is represented as a functional of the deformation field, it can be written in the Eulerian frame by inserting the inverse of the transformation $f^i$. Note that $\initial J$ can be set to $1$ if we require $x^i = X^i$ at initial time.
The closed Lagrange--Newton system is defined by the nonlinear gravitational evolution equations \eqref{curl_lagr_form} and \eqref{div_lagr_form} for the deformation gradient, see Ref. \cite{ehlersbuchert}:
\begin{eqnarray}
	\label{curl_lagr_form} \delta_{ij} \bd \ddot{f}^i \wedge \bd f^j & = & {\bf 0} \;, \\
	\label{div_lagr_form} \frac{1}{2} \epsilon_{ijk} \bd \ddot{f}^i \wedge \bd f^j \wedge \bd f^k & = & \left( \Lambda - 4 \pi G \varrho \right) \bd^3 f \;,
\end{eqnarray}
with $\Lambda$ the cosmological constant, $G$ the gravitational constant, 
Eq.~\eqref{curl_lagr_form} corresponding to the three field equations for the Eulerian curl, and Eq.~\eqref{div_lagr_form} corresponding to the field equation for the Eulerian divergence of the gravitational field strength. In the above equations the exact integral for the density (third equation of Eq.~(\ref{eulerfields})) has to be inserted to reduce the number of variables. For $\Lambda = 0$ the system does not explicitly contain the Jacobian, provided $J \ne0$. Regular solutions are characterized by $J >0$. 
Since, in the Newtonian theory, the Cartan coframe fields are exact forms, Eq.~(\ref{J}) reads:
\begin{eqnarray}
\bJ :=   \frac{\epsilon_{ijk}}{6}  \bd f^i \wedge \bd f^j \wedge \bd f^k  & = & \frac{\epsilon_{ijk}}{6} J \bd X^i \wedge \bd X^j \wedge \bd X^k
\non
    \Longleftrightarrow \bJ : = J \bd^3 X & = & \bd^3 f ;\; J = \det (f^k_{\;|i})\;\;,\quad\quad
\end{eqnarray}
where $\bd^3X$ is the Lagrangian volume $3-$form. The coefficients of these equations are equivalent to the following coefficient equations for the deformation gradient $\bd f^i$ in Lagrangian coordinates:
\begin{eqnarray}
	\label{curl_lagr} \delta_{ij} \ddot{f}^i_{\ |[p} f^j_{\ |q]} & = & 0 \;, \\
	\label{div_lagr} \frac{1}{2} \epsilon_{ijk} \epsilon^{pqr} \ddot{f} \indices{^i_{|p}} f \indices{^j_{|q}} f \indices{^k_{|r}} & = & \Lambda J - 4 \pi G \varrho J \;.
\end{eqnarray}
An alternative to express Eq.~\eqref{curl_lagr} reads: 
\begin{equation}
\label{curl2}
\delta^{kr} \epsilon_{pq[j} \epsilon^{ilm} \ddot{f}_ {r]|i} f \indices{^p_{|l}} f \indices{^q_{|m}} = 0\;.
\end{equation}
Expressed in terms of the Newtonian tidal tensor,
\begin{eqnarray}
&{\cal E}^i_{\ j} = \frac{1}{2J} \epsilon_{abc} \epsilon^{ikl} \ddot{f}^a_{\ | j} f^b_{\ |k} f^c_{\ | l} - \frac{1}{3}\left( \Lambda  - 4 \pi G \varrho \right)\delta^i_{\ j} \; ;\label{tidal}\\
&{\cal E}_{[ij]} = 0 \quad ; \quad {\cal E}^k_{\ k} = 0 \;,\label{ne_sys}
\end{eqnarray}
Eqs.~\eqref{ne_sys} correspond to the Lagrange--Newton system of equations $\lbrace\eqref{curl_lagr}, \eqref{div_lagr}\rbrace$.


\subsection{Einstein's Equations in Lagrangian form}
\label{sec:einsteinfull}
We formulate here the Einstein equations in terms of Cartan coframe fields as they are transported along the flow lines (here spacetime geodesics). The system of equations we obtain will be called the  Lagrange--Einstein system. 

In terms of coframe fields, the irrotational dust continuum is governed by the following evolution and constraint equations:
\begin{align}
\label{form_sym}&G_{ab}\, \ddot{\bet}^a \wedge \bet^b = {\bf 0} \;; \\
\label{form_evo}&\frac{1}{2} \epsilon_{dbc} \left( \dot{\bet}^a \wedge \bet^b \wedge \bet^c \right) \dot{} \non
&= \left( - \CR^a_{\ d} + \left( 4 \pi G \varrho + \Lambda \right) \delta^a_{\ d} \right) J \bd^3 X \; ;\\
\label{form_ham}&\epsilon_{abc} \,\dot{\bet}^a \wedge \dot{\bet}^b \wedge \bet^c = \left( 16 \pi G \varrho + 2 \Lambda - \CR \right) J \bd^3 X \; ; \\
\label{form_mom}&\epsilon_{abc} \, \left( \bd \dot{\bet}^a \wedge \bet^b + \boldsymbol{\omega}^a_{\ d} \wedge \dot{\bet}^d \wedge \bet^b \right) = {\bf 0} \;,
\end{align}
where the equations are, respectively, the irrotationality condition on the gravitational field \eqref{form_sym}, the equation of motion \eqref{form_evo}, the energy constraint \eqref{form_ham} and the momentum constraints \eqref{form_mom}. 
For irrotational matter flows, as is assumed throughout this paper, the first equation can be replaced by the kinematical irrotationality condition: 
\begin{equation}
\label{irrotationality}
G_{ab}\, \dot{\bet}^a \wedge \bet^b = {\bf 0}\;\;. 
\end{equation}
Nevertheless, we consider the double time--derivative expression for two reasons: (i) in the Newtonian limit we want to reproduce the field equations, which involve a second time--derivative, and (ii) for a general system, this equation is always true because of the conservation of the vorticity $2-$form, $\boldsymbol{\omega} = G_{ab} \,\dot{\bet}^a \wedge \bet^b = \initial {\boldsymbol{\omega}}$; for the Newtonian case, $\boldsymbol{\omega} =\bd \left( v_i \bd X^i \right) = \initial {\boldsymbol{\omega}}$ (see Appendix B in Ref. \cite{ehlersbuchert}). Nevertheless, it is important to note that the $3+1$ foliation formalism cannot describe a nonzero vorticity. The description of a nonzero vorticity will require a $1+3$ threading of spacetime (see, e.g., Ref. \cite{gourgoulhon07}) .

The combination of the trace of the equation of motion and the energy constraint straightforwardly leads to the Raychaudhuri equation:
\begin{eqnarray}
\label{Ray} \frac{1}{2} \epsilon_{abc} \ddot{\bet}^a \wedge \bet^b \wedge \bet^c = \left( \Lambda - 4 \pi G \varrho \right) J \bd^3 X\;.
\end{eqnarray}
To derive the above equations we have implicitly used the Cartan connection one--form and the curvature two--form that we do not need explicitly in what follows:
\begin{eqnarray}
\boldsymbol{\omega}^a_{\ b} := \gamma^a_{\ cb} \bet^c, \\
\mathbf{\Omega}^a_{\ b} := \frac{1}{2} \CR^a_{\ bcd} \bet^c \wedge \bet^d \;,
\end{eqnarray}
with the connection and curvature coefficients $\gamma^a_{\ cb}$ and $\CR^a_{\ bcd}$ in the nonexact basis, respectively. 
The 3--Ricci tensor can be expressed through the curvature two--form:
\begin{eqnarray}
\CR^a_{\ d} \bet^d \wedge \bet^b \wedge \bet^c = \delta^{db} \mathbf{\Omega}^a_{\ d} \wedge \bet^c - \delta^{dc} \mathbf{\Omega}^a_{\ d} \wedge \bet^b \;.
\end{eqnarray}
We employ the Hodge star operation to obtain the coefficient equations in the exact basis $\bd X^i$: 
\begin{align}
\label{form_symcoeff}&G_{ab} \,\ddot{\eta}^a_{[i} \eta^b_{\ j]} = 0 \;; \\
\label{form_eomcoeff}&\frac{1}{2 J} \epsilon_{abc} \epsilon^{ikl} \left( \dot{\eta}^a_{\ j} \eta^b_{\ k} \eta^c_{\ l} \right) \dot{} = -\CR^i_{\ j} + \left( 4 \pi G \varrho + \Lambda \right) \delta^i_{\ j}\;; \\
&\frac{1}{2J}\epsilon_{abc} \epsilon^{mjk} \dot{\eta}^a_{\ m} \dot{\eta}^b_{\ j} \eta^c_{\ k} = - \frac{\CR}{2}+ \left( 8\pi G \varrho +  \Lambda  \right) \;; \\
\label{form_momcoeff}&\left(\epsilon_{abc} \epsilon^{ikl} \dot{\eta}^a_{\ j} \eta^b_{\ k} \eta^c_{\ l} \right)_{||i} = \left(\epsilon_{abc} \epsilon^{ikl} \dot{\eta}^a_{\ i} \eta^b_{\ k} \eta^c_{\ l} \right)_{|j} \;,
\end{align}
where a double vertical slash denotes the covariant spatial derivative with respect to the $3$--metric and the spatial connection is assumed symmetric.
As before, the first equation can be replaced by the irrotationality condition: 
\begin{equation}
\label{irrotationality2}
G_{ab} \,\dot{\eta}^a_{[i} \eta^b_{\ j]} = 0 \;. 
\end{equation}
Again, the trace of the equation of motion and the energy constraint leads to the Raychaudhuri equation:
\begin{eqnarray}
\label{Raycoeff}\frac{1}{2J} \epsilon_{abc}\epsilon^{ik\ell} \ddot{\eta}^a_{\ i} \eta^b_{\ k} \eta^c_{\ \ell}   = \Lambda  - 4 \pi G {\varrho} \;.
\end{eqnarray}
\indent
The system $\lbrace\eqref{form_symcoeff}-\eqref{form_momcoeff}\rbrace$ consists of $13$ equations, where the first corresponds to the irrotationality condition ($3$ equations), the second to the symmetric evolution equations ($6$ equations), subjected to $4$ constraint equations that are the ADM constraints ($1$ equation for the energy constraint, and $3$ equations for the momentum constraints).  Thus, the first $9$ equations furnish evolution equations for the $9$ coefficient functions of the $3$ Cartan coframe fields.\par

The above system is equivalent to the results developed in Ref. \cite{rza1} in a different basis: in the first paper the choice of the standard orthonormal coframes has been made, whereas since Ref. \cite{rza2} the choice of the adapted coframes is preferred for reasons of allowing us to construct a formally closer Newtonian analogy.


\subsection{Equivalence of the two gravitoelectric sets of equations in the Minkowski Restriction}

We will now discuss the link between parts of the relativistic system and the full Newtonian system. Formally, this link is provided by the {\it Minkowski Restriction}.

\subsubsection{Definition of the Minkowski Restriction}
\label{subsec:MR}

Let $\bet^\alpha$ be Cartan one--form fields in a $4-$dimensional manifold (Greek letters are used in $4$ dimensions). A set of forms $\bet^\alpha$ is said to be {\it exact}, if there exist functions $f^\alpha$ such that $\bet^\alpha = \bd f^\alpha$, where $\bd$ denotes the exterior derivative operator, acting on forms and functions. The {\it Minkowski Restriction} (henceforth \MRP) consists of the replacement of the nonintegrable coefficients by integrable ones, $\eta^{\alpha}_{\ \nu} \rightarrow f^{\alpha \rightarrow \mu}_{\ | \nu}$, keeping the speed of light $c$ finite. With this restriction, the Cartan coframe coefficients yield the Newtonian deformation gradient, and the local tangent spaces all become identical and form the global Minkowski spacetime. The {\it Newtonian limit} could be defined as the \MR of Einstein's theory and additionally sending $c$ to infinity. In the flow--orthogonal foliation, employed in this paper, the $4-$dimensional coframes reduce to $\bet^\alpha =  (\bd t, \bet^a)$, and their \MR reads $\bd f^{\alpha} = ( \bd t, \bd f^{a \rightarrow i} )$. Note that $c$ and the signature are carried by the $4-$dimensional metric coefficients; $c$ is set to $1$ throughout this paper. We will need the inverse \MR and use it as a rule to construct relativistic Lagrangian solutions from known Newtonian solutions. (For the \MR of the metric, compare appendix~\ref{sec:appA}, and the remark on the Newton--GR ``dictionary'' in appendix~\ref{sec:appB}).

\subsubsection{Gravitoelectric equations}

Reference \cite{rza1} noted that a part of Einstein's equations, namely $\lbrace\eqref{form_sym}, \eqref{Ray}\rbrace$, are related to the gravitoelectric part of the spatially projected Weyl tensor. This tensor is trace--free and represented by the $3$ one--form fields ${\bf E}^a$ (see Ref. \cite{rza1} Eq.~(A23)):
\begin{equation}
{\bf E}^a = - \ddot{\bet}^a + \frac{1}{3} (\Lambda - 4\pi G \varrho )\bet^a \;;
\end{equation}
then, the irrotationality condition \eqref{form_sym} and the trace equation of motion \eqref{Ray} are generated by
\begin{eqnarray}
\label{E_eq} &G_{ab}\, {\bf E}^a \wedge \bet^b = {\bf 0} \quad;\quad \epsilon_{abc}\,{\bf E}^a \wedge \bet^b \wedge \bet^c = {\bf 0}\;.
\end{eqnarray}
These two equations are therefore referred to as the {\it gravitoelectric part} of Einstein's equations.  A projection of the gravitoelectric one--form fields and Equations \eqref{E_eq}, using the Hodge star operator, yields to their coefficient representation:
\begin{eqnarray}
&E^i_{\ j} = -\frac{1}{2J} \epsilon_{abc} \epsilon^{ikl} \ddot{\eta}^a_{\ j} \eta^b_{\ k} \eta^c_{\ l} + \frac{1}{3}\left( \Lambda  - 4 \pi G \varrho \right)\delta^i_{\ j} \; ;\label{ge}\\
&E_{[ij]} = 0 \quad ; \quad E^k_{\ k} = 0 \;.\label{ge_sys}
\end{eqnarray}
(Note that $E_{i j}:=\delta^b_{\ i}G_{b a} E^a_{\ j} = G_{a i}E^a_{\ j}$.) This just provides a rewriting of $\lbrace\eqref{form_symcoeff}, \eqref{Raycoeff}\rbrace$.
(A remark on the gravitomagnetic part of the spatially projected Weyl tensor can be found in appendix~\ref{sec:appB}.)

\subsubsection{Executing the \MR }
\label{MRexecute}

Sending the spatial Cartan coframes to exact forms, i.e., executing the \MRP,  their coefficients $\eta^a_{\ i}$ are restricted to the Newtonian deformation gradient $f^a_{\ |i}$. The gravitoelectric system of equations in the form of Eqs.~\eqref{ge_sys} then reduces to the Newtonian system in the form of Eqs.~\eqref{ne_sys}; note the conventional sign change between the gravitoelectric part of the spatially projected Weyl tensor $E^i_{\ j}$ and the Newtonian tidal tensor ${\cal E}^i_{\ j}$.
This operation closes the system, reducing the number of free functions from nine ($\eta^a_{\ i} (X^k ,t)$) to three ($f^i (X^k ,t)$). A consideration of the \MR for the remaining equations, yielding nontrivial Newtonian analogs, will not be needed in this paper, but will be the subject of forthcoming work.

Considering only the gravitoelectric equations is not enough to determine the nine functions of the coframe coefficients. The relativistic aspects contained in the remaining gravitomagnetic 
 equations will lead to a richer structure of the solutions and also to constraints on solutions of the gravitoelectric system. A followup work will explicitly consider both parts in the framework of first--order solutions. 
 
To conclude: the Lagrange--Einstein gravitoelectric equations are (up to nonintegrability) equivalent to their Newtonian analogs, whereas the gravitomagnetic equations have no obvious Newtonian counterpart (this issue is more subtle than the mere absence of a gravitomagnetic analogy, as we explain in the followup paper).


\section{Construction schemes for relativistic perturbations and solutions at any order}
\label{sec:n-order}

We now turn to the main part of this paper and construct the gravitoelectric subclass of  $n$th--order relativistic perturbation and solution schemes through generalization of the known Newtonian schemes.
This allows furnishing relativistic inhomogeneous models for large--scale structure formation in the Universe. The successful Lagrangian perturbation theory in Newtonian cosmology is well--developed. We will generalize here the perturbation and solution schemes of Newtonian cosmology given in the review \cite{ehlersbuchert}, the essential steps of which will be recalled in this section, followed by their relativistic counterparts.

All schemes are applied to the matter model `irrotational dust'. It is possible to extend the present schemes by employing the framework for more general fluids in a Lagrangian description that will be developed in forthcoming work. Most of the known representations are focused on writing equations in terms of tensor or form coefficients. Our investigation will be guided by the compact differential forms formalism as before. However, we will also project to the coefficient form in parallel to ease reading.

\subsection{General $\boldsymbol{\it n}$th--order perturbation scheme}	

As in standard perturbation theories, we decompose the perturbed quantity into a Friedmann--Lema\^\i tre--Robertson--Walker (FLRW) solution and deviations thereof, which are expanded up to a chosen order $n$ of the perturbations. Contrary to the standard perturbation theory, we do not perturb the metric globally at the background space, but we perturb the Cartan coframes locally:
\begin{eqnarray}
 \bet^a = \eta^a_{\ i} \bd X^i = a(t) \left( \delta^a_{\ i} + \displaystyle\sum_n P^{a \sm{(n)}}_{\ i} \right) \bd X^i \;,%
\label{pert_scheme}
\end{eqnarray}
in the local exact basis $\bd X^i$. Notice that with this ansatz we choose to perturb a zero--curvature FLRW model, but it is possible to encode an initial first--order constant curvature in the coefficient functions $G_{ab}$ in the following local metric coefficients, which can be calculated from the above coframe ansatz:
\begin{eqnarray}
g_{ij} = G_{ab} \eta^{a}_{\ i} \eta^{b}_{\ j} \;\;.
\end{eqnarray}
Furthermore, we can link these results to the ones obtained for the orthonormal coframes $\tilde{\boldsymbol{\eta}}^c$ (compare also the corresponding remarks in \cite{chandra}:~p.35 and \cite{gron}:~p.75). Indeed, the metric bilinear form can be written as:
\begin{equation}
\mathbf{g} = \delta_{cd} \ \tilde{\boldsymbol{\eta}}^c \otimes \tilde{\boldsymbol{\eta}}^d = G_{ab} \ {\boldsymbol{\eta}}^a \otimes {\boldsymbol{\eta}}^b \:\:.
\label{apA}
\end{equation}
From this identity, we conclude:
\begin{equation}
G_{ab} = \delta_{cd} \ \tilde{{\eta}}^c_{\ a}\tilde{{\eta}}^d_{\ b} \:\:,
\label{Gabdelta}
\end{equation}
where the $\tilde{{\eta}}^c_{\ a}$ are the coefficients of the projection of $\tilde{\boldsymbol{\eta}}^c$ onto the basis ${\boldsymbol{\eta}}^a$.
In the next subsection, we will specify the coframes we consider in such a way that the initial coframe perturbations vanish. From now on, we will call these coframes \textsl{adapted coframes} to distinguish them from the orthonormal ones (see also appendix~\ref{sec:appA}).

	\subsection{Initial data for the perturbation scheme}

We choose initial data in formal correspondence with the Lagrangian theory in Newtonian cosmology and generalize these initial fields to the relativistic stage. This has obvious advantages with regard to the aim to give construction rules that translate the known Newtonian solutions to general relativity. For the initial data setting in the Newtonian case, see Ref. \cite{ehlersbuchert}.

		\subsubsection{Fundamental initial data}	
		
Let the three one--form fields $\mathbf{U}^a = U^a_{\ i} \bd X^i$ be the initial one--form generalization of the Newtonian peculiar velocity--gradient, obtained by the inverse \MRP. Accordingly, let $\mathbf{W}^a = W^a_{\ i} \bd X^i$ be the initial one--form generalization of the Newtonian peculiar--acceleration gradient. Our solutions will be written in terms of these initial data. They determine the initial values of the coframes as follows:
\begin{eqnarray}
&&\eta^a_{\ i}(\initial t) = \delta^a_{\ i} \;, \label{prem}\\
&&\dot{\eta}^a_{\ i}(\initial t) = \initial H \delta^a_{\ i} + U^a_{\ i} \;;\; \initial H : = \initial{\dot{a}} \;;\; \initial{a} : =1 \;; \label{inipec1}\\ 
&&\ddot{\eta}^a_{\ i}(\initial t) = \initial{\ddot{a}} \delta^a_{\ i} + W^a_{\ i} \label{inipec2}\;.
\end{eqnarray}
Equation~\eqref{prem} implies that the coframes we will work with from now on are initially equal to the exact Lagrangian coordinate basis: $\boldsymbol{\eta}^a(\initial t) = \delta^a_{\ i} \ \mathbf{d} X^i$.
This in turn provides the initial metric coefficients in the form:
\begin{eqnarray}
G_{ij} =g_{ij}(\initial t) = G_{ab} \delta^a_{\ i} \delta^b_{\ j}  \;\;.
\end{eqnarray}

In view of the flow--orthogonal foliation, we have the irrotationality constraint:
\begin{equation}
\boldsymbol{\omega} = G_{ab} \dot{\bet}^a \wedge \bet^b = {\bf 0} \; \Longrightarrow \; G_{ab} \mathbf{U}^a \wedge \delta^b_{\ j} \bd X^j= {\bf 0} \;.
\end{equation}
This implies for the coefficient functions: $U_{[ij]} = 0$. (We used the implicit definition $U_{ij} : = \delta^b_{\ i}G_{b a} U^a_{\ j}$.)\\

\noindent
{\it Remark:}

\noindent
From \eqref{Gabdelta} and \eqref{prem}, it is interesting to notice the following relations that hold to zeroth and first order (the full initial data are considered to be first order, as was also the choice in the Newtonian schemes \cite{ehlersbuchert}):
\begin{equation}
    \left\{
    \begin{split}
    G^{\sm{(0)}}_{ij} &= \delta_{ij} \;; \\
    G^{\sm{(1)}}_{ij} &= 2 \ \hspace{-.1cm} \tilde{\CP}_{ij} \;; \\
     \label{plorky}2\tilde{P}_{(ij)} &= G^{\sm{(1)}}_{ij} +2 {P}_{(ij)}\;;\\
    \end{split}
    \right.
\end{equation} 
where $\tilde{\CP}_{ij} =\tilde{P}_{ij}(\initial t)$. We are thus able to rederive some results from the ones obtained in previous works that used orthonormal coframes. For example, the Ricci curvature tensor at first--order can be obtained by injecting the identities \eqref{plorky} into (93) of Ref. \cite{rza1}. We can so obtain the adapted coframes from the orthonormal ones and vice--versa (compare also appendix~\ref{sec:appA}).

\subsubsection{Relativistic counterpart of the Poisson equation and consequences for $ \mathbf{W}^a $}
\label{relpois}

In the Newtonian approach the initial peculiar--acceleration and the density inhomogeneities are linked through the Poisson equation. To generalize this equation to the relativistic case, we note the relativistic generalization of the Newtonian field strength gradient that follows from inspection of the Lagrange--Einstein system (for details the reader can always consult Ref.~\cite{rza1}):
\begin{eqnarray}
\label{genefieldstren}
\CF^i_{\ j}  :=  \dot{\Theta}^i_{\ j} + \Theta^i_{\ k} \Theta^k_{\ j}\qquad\qquad \nonumber\\ 
 = - \CR^i_{\ j} - \Theta \Theta^i_{\ j} + \left( 4 \pi G \varrho - \Lambda \right) \delta^i_{\ j} + \Theta^i_{\ k} \Theta^k_{\ j}\;,
\end{eqnarray}
with the $3-$Ricci tensor coefficients $\CR_{ij}$ of which the trace is the Ricci scalar $\CR$, and $\Theta_{i j}$ the expansion tensor coefficients. According to the energy constraint, $\CR + \Theta^2 - \Theta^{k}_{\ \ell} \Theta^{\ell}_{\ k}= 
6 \pi G \varrho + 2 \Lambda$, the symmetry of the expansion tensor and Ricci curvature, it is straightforward to show that the relativistic gravitational field coefficients $\CF_{i j}$ respect the following field equations:
\begin{equation}
    \CF^k_{\ k} = \Lambda - 4 \pi G \varrho  \quad;\quad \CF_{[ij]} = 0 \;.
\end{equation}
In terms of the coframe fields, the relativistic gravitational field can be written as follows:
\begin{equation}
\label{GRgrav}
\CF^i_{\ j}  =  \frac{1}{2J} \epsilon_{abc} \epsilon^{ikl} {\ddot{\eta}}^{a}_{\ j} {\eta}^{b}_{\ k} {\eta}^{c}_{\ l} \;\;.
\end{equation}
(For an alternative derivation using a Newton--GR ``dictionary'' see appendix~\ref{sec:appB}).

Hence, inserting the coframe perturbations and evaluating this expression at initial time, we get the following relations (note that the zero--order fields trivially satisfy the second constraint):
\begin{eqnarray}
    \CF^k_{\ k}(\initial t) = \Lambda - 4 \pi G \initial \varrho = \qquad\qquad\qquad\qquad \non
    \Lambda - 4\pi G\varrho_{H \rm \bf i}(1+ \initial\delta ) = 3 \initial{\ddot{a}} + \delta^k_{\ a } W^a_{\ k} \;; \non
    \CF_{[ij]}(\initial t) = \delta^b_{\ [i}G_{b a} W^a_{\ j]} = W_{[ij]} = 0 \;.\qquad\qquad ,
\end{eqnarray} 
with the initial density contrast $\initial\delta$.
Thus, the deviation one--form fields $\bW^a$ obey the following equations that generalize the Poisson equation for the inhomogeneous deviations off the zero--order solution:
\begin{eqnarray}
\ast \frac{1}{2} \epsilon_{abc} \mathbf{W}^a \wedge \delta^b_{\ j} \bd X^j \wedge \delta^c_{\ k} \bd X^k =  - 4 \pi G \delta \initial \varrho \;;\;  \non
G_{ab} \mathbf{W}^a \wedge \delta^b_{\ j} \bd X^j = \mathbf{0}\;,\qquad\qquad
\label{eq:genW}
\end{eqnarray}
with $\delta \initial \varrho = \initial \varrho -  \varrho_{H \rm \bf i}$, implying for the coefficient functions:
\begin{eqnarray}
- \frac{1 }{4\pi G} \delta^k_{\ a } W^a_{\ k} &=& \delta \initial \varrho = \varrho_{H \rm \bf i} \initial \delta \quad;\quad W_{[ij]} = 0\;.
\label{eq:genpoiss}
\end{eqnarray}
\subsubsection{Summary of initial data}
\label{initialdata}

We summarize the set of initial data, determined by our choice of the basis and subjected to the constraints. We assume in perturbative expansions, without loss of generality \cite{ehlersbuchert}, that the initial data $\lbrace\eqref{initialdata1}-\eqref{initialdata3}\rbrace$ are first order. We drop the index ${}^{\sm{(1)}}$ for notational ease and denote the initial data for the comoving perturbation form coefficients by $P^{a}_{\ i} (t_{i}) =: \CP^a_{\ i}$. We set:
\begin{itemize}
		\item for the initial deformation and the initial generalizations of the Newtonian velocity and acceleration gradients:
		\begin{align}
		  \begin{cases}
		    \label{initialdata1}
		   \boldsymbol{\CP}^{a \sm{(n)}} = \mathbf{0} \; \forall n \;, \\
		   \mathbf{U}^{a \sm{(1)}} = \mathbf{U}^a, \; U_{[ij]} = 0 \;; \\
		    \mathbf{W}^{a \sm{(1)}} = \mathbf{W}^a, \; W_{[ij]} = 0 \;,
		  \end{cases}
		\end{align}
		\item where the coefficients are related via the initial values of the time--derivatives of the deformation:
		\begin{align}
		  \begin{cases}
		    \label{initialdata2}
		\dot{\CP}^a_{\ i} = U^a_{\ i} \;; \\ 
		\ddot{\CP}^a_{\ i} = W^a_{\ i} - 2 \initial H U^a_{\ i} \;, 
		  \end{cases}
		\end{align}  
		\item together with additional initial constraints that are to be respected (a relation to the initial metric, to the initial density contrast, and the four ADM constraint equations evaluated at initial time):
		\begin{align}
		  \begin{cases}
		    \label{initialdata3}
		G_{ij} = G_{ab} \,\eta^a_{\ i} (\initial t) \eta^b_{\ j} (\initial t) = G_{ab}\delta^a_{\ i}\delta^b_{\ j}\;;\;\\
		 4\pi G \delta \initial \varrho^{\sm{(1)}} = - W\;;\\
    		\initial H U =-\frac{\CR (\initial t)}{4} - W \;;\\
	    	\left(U_{\; j}^{a}\delta_{a}^{\; i}\right)_{||i}=\left(U_{\; i}^{a}\delta_{a}^{\; i}\right)_{|j} \;;
	    	  \end{cases}
		\end{align}
(here and in the following we use the abbreviations $\delta^k_{\ a } U^a_{\ k} = :U$, $\delta^k_{\ a } W^a_{\ k} = :W$ for the trace expressions.)	
		\item the initial Ricci curvature as found from the equation of motion \eqref{form_eomcoeff}:
		\begin{eqnarray}
		\label{kinematicRicci}
		 \CR^{i}_{j}(\initial t) =  - (W^{i}_{j} +\initial H U^{i}_{j} ) - \delta^{i}_{\ j} (W +\initial H U) \nonumber \\ -\epsilon_{abc}\epsilon^{ilk}U_{\ j}^{a}U_{\ l}^{b}\delta_{\ k}^{c}\:\;;
		\end{eqnarray}
					
		\item equating this expression with the initial Ricci tensor as calculated from the initial metric,
\begin{align}
\label{geometricalRicci}\qquad
&{\CR}_{\ j}^{i} (\initial t) = G_{[j|b]}^{\, i\sm{(1)}} G_{a}^{\, b|a\sm{(1)}} + G^{\, b|a\sm{(1)}}_{j} G_{[b|a]}^{\, i\sm{(1)}}\non
&+ G_{b|j}^{\, a\sm{(1)}} G_{a}^{\,[b|i]\sm{(1)}}+\frac{1}{2}\, G^{\, a\sm{(1)}}_{[a|b]} G_{j}^{\, b|i\sm{(1)}}
- \frac{1}{2} G^{\, a|b\sm{(1)}}_{a}\, G^{\, i\sm{(1)}}_{[j|b]}\non 
&+ \frac{1}{2} G_{b|[j}^{{\sm{(1)}}\, a} G^{{\sm{(1)}}\, b|i}_{a]} + 2 G_{[j|a]}^{{\sm{(1)}}\,[a|i]} + 2 G_{[j|a]}^{{\sm{(2)}}\,[a|i]}\non 
&- 2 G_{\,b}^{{\sm{(1)}}\, a} G^{[b|i]\sm{(1)}}_{[j|a]}-2G_{a}^{{\sm{(1)}}\, i}G^{[b|a]\sm{(1)}}_{[j|b]},
\end{align}
where, e.g., $G_{ia} = G_{ij}\delta^{j}_{\ a}$ and $G^i_{\ j} = G_{k j}\delta^{ki}$,
we determine the first--order part of the initial metric (which is a derived quantity),
		\begin{equation}
		\label{G1eq} \qquad\,
		 2G_{\;[k|j]}^{{\sm{(1)}}\,[i|k]}=-\initial HU_{\ j}^{i}-W_{\ j}^{i}-\left(\initial HU+W\right)\delta_{\ j}^{i}\:\:,
		\end{equation}
as well as the second--order part of the initial metric (which later appears in the perturbation and solution schemes):
		\begin{equation}
		\label{G2eq}
		 2G_{\;[k|j]}^{{\sm{(2)}}\,[i|k]}=f(U_{\ j}^{i},W_{\ j}^{i})\:\:,
		\end{equation}
\end{itemize}
where the function $f$ can again be derived by equating \eqref{geometricalRicci} and \eqref{kinematicRicci}. All further order right--hand sides of \eqref{kinematicRicci} vanish.

The initial data given in \eqref{initialdata1} are exhaustive: in our ADM split, the system of equations $\lbrace\eqref{form_symcoeff}-\eqref{form_momcoeff}\rbrace$ contains $9$ second--order equations of motion for the coframes  subjected to $4$ constraints. A general solution therefore contains $18$ coefficient functions encoded in $U_{ij}$ and $W_{ij}$ that reduce to $12$ functions for solutions of the irrotationality conditions \eqref{form_symcoeff}, the latter of which are represented by the $6$ constraints $U_{[ij]}=0$ and $W_{[ij]}=0$. The general solution is further subjected to the $4$ ADM constraints resulting in corresponding constraints on $U_{ij}$ and $W_{ij}$.

\subsection{Gravitoelectric Perturbation Scheme}

We now recall the general Lagrangian perturbation scheme of Newtonian cosmology and generalize it to a gravitoelectric scheme in relativistic cosmology. By construction, this latter will already contain the known Lagrangian perturbation scheme at any order in the geometrical limit of exact deformation one--forms.

		\subsubsection{Recap: Newtonian Theory}
\label{sec:newtonperturbations}
The general perturbation scheme has been fully developed in Ref. \cite{ehlersbuchert}. Our approach only slightly differs in terms of the initial conditions: we formulate them such that they are formally closer to the relativistic approach. Following the general ansatz \eqref{pert_scheme}, we introduce three comoving perturbation forms $\bd P^i$ of the three components of the comoving vector perturbation fields $P^i (X^i,t)$: 
\begin{equation}
\bd f^i ( \vec{X},t ) = :  a(t) \bd F^i ( \vec{X},t ) =  a(t) \left( \bd X^i + \bd P^i ( \vec{X},t) \right) \;,
\label{pert_newt}
\end{equation}
and decompose the perturbation gradient field on the FLRW background order by order:
\begin{equation}
    \bd P^i =  \displaystyle\sum_{m=1}^{\infty} \epsilon^m \bd P^{i \sm{(m)}} \;.
     \label{pert_decomp}
\end{equation}
It is, of course, possible to consider perturbations of the position fields $f^i$, because the Newtonian equation can be expressed in a vectorial form. The relativistic equations are, however, tensorial, and we therefore consider the representation in terms of the gradient of the fluid's deformation. 

To provide unique solutions of the Newtonian system, suitable boundary conditions have to be imposed. For the cosmological framework the requirement of periodic boundary conditions for field deviations from a Hubble flow is a possible choice \cite{buchertehlers:average}. This translates into an integral constraint on the perturbations: integration over a compact spatial domain $\cal M$ implies the following: 
\begin{equation}
\int \limits_{\cal M} \bd P^i = \displaystyle\int \limits_{\partial{\cal M}} P^i =0 \quad;\quad   P^i =  \displaystyle\sum_{m=1}^{\infty} \epsilon^m P^{i \sm{(m)}}.
\end{equation}
Recall now that $\mathbf{U}^i = \bd U^i = U^i_{\ |j} \bd X^j$ and $\mathbf{W}^i = \bd W^i = W^i_{\ |j} \bd X^j$ are the initial one--form peculiar--velocity gradient and the initial one--form peculiar--acceleration gradient. 
The fields ${W}^i$ are determined nonlocally by the following set of equations, equivalent to Poisson's equation:
\begin{eqnarray}
W^i_{\ |i} = \ast \frac{1}{2} \epsilon_{ijk} \bd W^i \wedge \bd X^j \wedge \bd X^k = - 4 \pi G \delta \initial\varrho\;;\; \non
\delta_{ij} \bd W^i \wedge \bd X^j = \ast \bd \left( W^i \bd X^i \right) = {\bf 0}\;.
\end{eqnarray} 
In view of the restriction to irrotational flows, we additionally impose the constraint: 
\begin{equation}
\delta_{ij} \bd \dot{f}^i \wedge \bd f^j = {\bf 0} \Longrightarrow \delta_{ij} \bd U^i \wedge \bd X^j = \ast \bd \left( U^i \bd X^i \right) = {\bf 0} \;.
\end{equation}
		Without loss of generality, we can choose the following general set of initial data that can be obtained in the Newtonian theory or else from the \MR
		of (\eqref{initialdata1}--\eqref{initialdata3}):
		\begin{itemize}
		\item for the initial deformation, peculiar--velocity and peculiar--acceleration:
		\begin{align}
		\label{MRinitialdata1}
		\begin{cases}
		 \bd \CP^{i \sm{(n)}} = \mathbf{0}\; \forall n\;;\\
		 \bd U^{i \sm{(1)}} = \bd U^i, \ U_{[i|j]} = 0 \;; \\		   
	         \bd W^{i \sm{(1)}} = \bd W^i, \ W_{[i|j]} = 0 \;,
		\end{cases}
		\end{align}   
		\item together with the definition of the Lagrangian metric coefficients and the initial data relation to the density perturbation:
		\begin{align}
		\label{MRinitialdata3}
		\begin{cases}
		g_{ij} = \delta_{kl} f^k_{\ |i} f^l_{\ |j} \;; \\
	         \delta\initial \varrho {} ^{\sm{(1)}} = \delta \initial\varrho = \varrho_{H \mathrm{ \mathbf{ i}}} \initial\delta = - \frac{1}{4\pi G} W^k_{\ | k} \;.
	    	\end{cases}
		\end{align}   
		\end{itemize}
The metric is Euclidean, since the coefficients can be transformed to the coefficients $\delta_{ij}$ with the help of the to $\vec f$ inverse coordinate transformation. 

Plugging the ansatz \eqref{pert_newt} into the Newtonian equations $\lbrace\eqref{curl_lagr_form}, \eqref{div_lagr_form}\rbrace$, we find for the background Friedmann's equation:
\begin{eqnarray}
	\label{newton_homog} &&\epsilon_{ijk} \ 3 \frac{\ddot{a}}{a} \bd X^i \wedge \bd X^j \wedge \bd X^k = \non
	&&\epsilon_{ijk} \left( \Lambda - 4 \pi G \varrho_H \right) \bd X^i \wedge \bd X^j \wedge \bd X^k \non
	&& \hspace{1cm} \Longrightarrow 3 \frac{\ddot{a}}{a} = \Lambda - 4 \pi G \varrho_H \;,
\end{eqnarray}
and a full hierarchy of the perturbation equations:
\begin{eqnarray}
	\label{curl_lagr_form_pert} & \delta_{ij} & \bd \dot{P}^i \wedge \left( \bd X^j +  \bd P^j \right) = \delta_{ij} a^{-2} \bd U^i \wedge \bd X^j \;; \\
	\label{div_lagr_form_pert} & \epsilon_{ijk} & \Bigg[ \left( {\CD_{1}} \bd P^i \right) \wedge \bd X^j \wedge \bd X^k
	 +  \left( 2 {\CD_{2}} \bd P^i \right) \wedge \bd P^j \wedge \bd X^k \non
	& + & \left( {\CD_{3}} \bd P^i \right) \wedge \bd P^j \wedge \bd P^k \Bigg] \non
	& = & - \epsilon_{ijk} \frac{4 \pi G}{3} \delta \initial \varrho a^{-3} \bd X^i \wedge \bd X^j \wedge \bd X^k \;,
\end{eqnarray}
where we defined the operator 
\begin{equation}
{\CD_{\ell}} := \frac{d^2}{dt^2} + 2 H \frac{d}{dt} - \frac{4}{\ell} \pi G \varrho_H\; .%
\label{def_Dx}
\end{equation}
(The reader may note a difference in the numerical coefficients from the Ref. \cite{ehlersbuchert}, which we had to correct, see appendix~\ref{sec:appB}.) Projecting with the Hodge star operator to the coefficient form (and integrating Eq.~\eqref{curl_lagr_form_pert}), we obtain:
\begin{eqnarray}
&&P_{[i|j]} = \displaystyle\int_{\initial t}^t \dot{P}_{m|[i} P^m_{\ |j]} dt' \;; \\
&&{\CD_{1}} P^i_{\ |i} = - 4 \pi G \delta \initial \varrho a^{-3} \label{trace_form_gen_pert} \\
&& -\frac{1}{2} \epsilon_{ijk} \epsilon^{lmn} \left[ P^i_{\ |l} \ P^j_{\ |m} {\CD_{3}} P^k_{\ |n} + 2 \delta^i_{\ |l} P^j_{\ |m} {\CD_{2}} P^k_{\ |n} \right]. \nonumber 
\end{eqnarray}
After splitting Eqs. $\eqref{curl_lagr_form_pert}$ and $\eqref{div_lagr_form_pert}$ order by order, we obtain $n$ sets of equations. 
At first--order we get:
\begin{eqnarray}
	\label{fcurl_1} \delta_{ij} \bd \dot{P}^{i \sm{(1)}} \wedge \bd X^j &=& \mathbf{0} \;;\qquad\\
	\label{fdiv_1}  \epsilon_{ijk} {\CD_{1}} \bd P^{i \sm{(1)}} \wedge \bd X^j \wedge \bd X^k &=&\non
	 a^{-3} \epsilon_{ijk} \bd W^i \wedge \bd X^j \wedge \bd X^k \;; 
\end{eqnarray}
in coefficient form:
\begin{eqnarray}
	P^{\sm{(1)}}_{[i|j]} = 0 \quad;\quad{\CD_{1}} P^{i \sm{(1)}}_{\ |i} = a^{-3} W^i_{\ |i} \;,
\end{eqnarray}
i.e., a set of linear equations. 
The generic $n$th--order system of equations will be written below with an implicit summation over the order of perturbations in the source terms:
\begin{eqnarray}
&& A^{\sm{(p)}} B^{\sm{(q)}} = \displaystyle\sum_{p+q=n} A^{\sm{(p)}} B^{\sm{(q)}} \;, \\
&& A^{\sm{(r)}} B^{\sm{(s)}} C^{\sm{(t)}} = \displaystyle\sum_{r+s+t=n} A^{\sm{(r)}} B^{\sm{(s)}} C^{\sm{(t)}} \;.
\end{eqnarray}
Thus, at any order $n>1$, the perturbation equations read:
\begin{eqnarray}
    \label{Nform_curl_pertm} \delta_{ij}\bd \dot{P}^{i \sm{(n)}} \wedge \bd X^j &=& -\delta_{ij} \bd \dot{P}^{i \sm{(p)}} \wedge \bd P^{j \sm{(q)}}  \;;\qquad \\
    \label{Nform_div_pertm} \epsilon_{ijk} {\CD_{1}} \bd P^{i \sm{(n)}} \wedge \bd X^j \wedge \bd X^k &=&\non
    && \hspace{-3cm} -\epsilon_{ijk} \Big[ \left( 2 {\CD_{2}} \bd P^{i \sm{(p)}}\right) \wedge \bd P^{j \sm{(q)}} \wedge \bd X^k \non
    && \hspace{-3cm} + \left( {\CD_{3}} \bd P^{i \sm{(r)}} \right) \wedge \bd P^{j \sm{(s)}} \wedge \bd P^{k \sm{(t)}} \Big] \;;
\end{eqnarray}
in coefficient form:
\begin{eqnarray}
    P^{\sm{(n)}}_{[i|j]} &=& \displaystyle\int_{\initial t}^t \dot{P}^{\sm{(p)}}_{m|[i} P^{m \sm{(q)}}_{\ |j]} dt' \;; \\
 {\CD_{1}} P^{i \sm{(n)}}_{\ |i} &=&  - \frac{1}{2} \epsilon_{ijk} \epsilon^{lmn} P^{j \sm{(s)}}_{\ |m} P^{k \sm{(t)}}_{\ |n} {\CD_{3}} P^{i \sm{(r)}}_{\ |l} \non
&&  - \left( {\CD_{2}} P^{i \sm{(p)}}_{\ |i} \right) P^{j \sm{(q)}}_{\ |j} + \left( {\CD_{2}} P^{i \sm{(p)}}_{\ |j} \right) P^{j \sm{(q)}}_{\ |i} .
\end{eqnarray}
The reader may consult the review \cite{ehlersbuchert} and references therein for further details.

		\subsubsection{Einstein's Theory}
\label{sec:generalperturbation}
Assuming the perturbation ansatz \eqref{pert_scheme} for the coframes, and using the operator ${\CD_{\ell}}$ as defined in \eqref{def_Dx}, the analogous expansion is performed: 
the zeroth order again leads to the Friedmann equation, and the general perturbation scheme reads:
\begin{eqnarray}
	\label{Einsteinsym_form_pert} & G_{ab} & \dot{\bP}^a \wedge \delta^b_{\ j} \bd X^j + G_{ab} \dot{\bP}^a \wedge \bP^b = \mathbf{0} \;;\\
	\label{Einsteintr_form_pert} & \epsilon_{abc} & \Bigg[{\CD_{1}}  \bP^a \wedge \delta^b_{\ j} \bd X^j \wedge \delta^c_{\ k} \bd X^k \non
	& + & (2 {\CD_{2}} \bP^a) \wedge \bP^b \wedge \delta^c_{\ k} \bd X^k + \left( {\CD_{3}} \bP^a \right) \wedge \bP^b \wedge \bP^c \Bigg]\non
	& = & \epsilon_{abc} W \frac{1}{3} a^{-3} \delta^a_{\ i} \bd X^i \wedge \delta^b_{\ j} \bd X^j \wedge \delta^c_{\ k} \bd X^k \;.\non
\end{eqnarray}
In coefficient form and integrating Eq. \eqref{Einsteinsym_form_pert} they become:
\begin{eqnarray}
&&P_{[ij]} = G_{a b}\displaystyle\int_{\initial t}^t \dot{P}^{a}_{[i} P^b_{\ j]} dt' \;;\\
&&{\CD_{1}} P^i_{\ i} = -  \left( \left({\CD_{2}} P^i_{\ i} \right) P^j_{\ j} - \left( {\CD_{2}} P^i_{\ j} \right) P^j_{\ i} \right) \non
&& -\frac{1}{2} \epsilon_{ijk} \epsilon^{lmn} \left( {\CD_{3}} P^i_{\ |l} \right) P^j_{\ |m} P^k_{\ |n} + W a^{-3}\;.
\end{eqnarray}
Expansion order by order leads to the first--order gravitoelectric equations:
\begin{eqnarray}
	\label{elec1_1} G^{\sm{(0)}}_{ab} \dot{\bP}^{a \sm{(1)}} \wedge \bd X^b = \mathbf{0} \;; \\
	\label{elec2_1} \epsilon_{abc} {\CD_{1}} \mathbf{P}^{a \sm{(1)}} \wedge \delta^b_{\ j} \bd X^j \wedge \delta^c_{\ k} \bd X^k \non
     =  a^{-3} \epsilon_{abc} \mathbf{W}^a \wedge \delta^b_{\ j} \bd X^j \wedge \delta^c_{\ k} \bd X^k \;;
\end{eqnarray}
and the general $n$th--order, $n>1$, set of nonlinear equations:
\begin{equation}
    \label{elec1_pert} G_{ab} \dot{\bP}^{a \sm{(n)}} \wedge \delta^b_{\ j} \bd X^j = - G_{ab} \dot{\bP}^{a \sm{(p)}} \wedge \bP^{b \sm{(q)}} \;;
    \end{equation}
    \begin{align}
    \label{elec2_pert} 
     \epsilon_{abc} & {\CD_{1}} \mathbf{P}^{a \sm{(n)}} \wedge \delta^b_{\ j} \bd X^j \wedge \delta^c_{\ k} \bd X^k = \nonumber \\
      -\epsilon_{abc} & \left[ 2  \left( {\CD_{2}} \mathbf{P}^{a \sm{(p)}} \right) \wedge \mathbf{P}^{b \sm{(q)}} \wedge \delta^c_{\ k} \bd X^k\right. \\ 
     & \; \left. + \left( {\CD_{3}} \mathbf{P}^{a \sm{(r)}} \right) \wedge \mathbf{P}^{b \sm{(s)}} \wedge \mathbf{P}^{c \sm{(t)}} \right].\no
   \end{align}
In coefficient form, this reads:
\begin{eqnarray}
\label{elec1_pert_coeff_1}
	P^{\sm{(1)}}_{[ij]} = 0 \quad ;\quad
    {\CD_{1}} P^{i \sm{(1)}}_{\ i} = W a^{-3}\; ,
\end{eqnarray}
and
\begin{eqnarray}
P^{\sm{(n)}}_{[ij]} &=& G^{\sm{(r)}}_{a b} \displaystyle\int_{\initial t}^t \dot{P}^{a \sm{(s)}}_{[i} P^{b \sm{(t)}}_{\ j]} dt' \;;
\label{eq:asysGR} \\
{\CD_{1}} P^{i \sm{(n)}}_{\ i} &=& - \frac{1}{2} \epsilon_{ijk} \epsilon^{lmn} P^{j \sm{(s)}}_{\ m} P^{k \sm{(t)}}_{\ n} \left( {\CD_{3}} P^{i \sm{(r)}}_{\ l} \right)  \non
 &&- \left( {\CD_{2}} P^{i \sm{(p)}}_{\ i} \right) P^{j \sm{(q)}}_{\ j}  + \left( {\CD_{2}} P^{i \sm{(p)}}_{\ j} \right) P^{j \sm{(q)}}_{\ i} .
\label{eq:trsysGR}
\end{eqnarray}
This provides equations for the perturbation fields at any order $n$ from solutions of order $n-1$. 

Comparing $\lbrace\eqref{elec1_pert}, \eqref{elec2_pert}\rbrace$ to the Newtonian equations $\lbrace\eqref{Nform_curl_pertm}, \eqref{Nform_div_pertm}\rbrace$, we see (not surprisingly) that we arrive at two equivalent sets of equations if we link the perturbations $\bP^a$ and $\bd P^i$ at any order via the \MR ---
recall that the construction was done by inversion of the \MRP, $\bd P^i = P^i_{\ |j} \bd X^j \mapsto P^a_{\ j} \bd X^j = \bP^a$; for the initial data: $\bd U^i = U^i_{\ |j} \bd X^j \mapsto U^a_{\ j} \bd X^j = \mathbf{U}^a$ and $\bd W^i = W^i_{\ |j} \bd X^j \mapsto W^a_{\ j} \bd X^j = \mathbf{W}^a$.
Therefore, we can simply translate the formal solution scheme for the trace--parts and the antisymmetric parts of the perturbations.
However, note already here that the inversion of the \MR produces a symmetric traceless component that is represented in Newtonian theory by the tidal tensor.
 
	\subsection{Gravitoelectric Solution Scheme}
\label{sec:gesolscheme}
		\subsubsection{Recap: Newtonian Theory}
\label{sec:newtonsolution}

We first recall the general solution scheme given in Ref. \cite{ehlersbuchert}, written for the perturbation gradients only. 

The hierarchy begins with the first--order equations $\lbrace\eqref{fcurl_1}, \eqref{fdiv_1}\rbrace$  which are uniquely determined by the constraint initial data \eqref{MRinitialdata1}.
The general $n$th--order, $n>1$, solution scheme from Eqs. $\lbrace\eqref{curl_lagr_form_pert}, \eqref{div_lagr_form_pert}\rbrace$, reads:
\begin{eqnarray}
   \label{curl_pertn}  \delta_{ij}  \bd P^{i \sm{(n)}} \wedge \bd X^j &=&  {}^N \boldsymbol{{\cal S}}^{\sm{(n)}}\;; \\
   \label{div_pertn} \epsilon_{ijk} {\CD_{1}} \bd P^{i \sm{(n)}} \wedge \bd X^j \wedge \bd X^k &=&  {}^N \boldsymbol{{\cal T}}^{\sm{(n)}},
\end{eqnarray}
uniquely determined by the source terms:
   \begin{eqnarray}
   \label{source_curl} {}^N \boldsymbol{{\cal S}}^{\sm{(n)}} & := & - \delta_{ij} \displaystyle\int_{t_0}^t \bd \dot{P}^{i \sm{(p)}} \wedge \bd P^{j \sm{(q)}} dt' \;; \\
   \label{source_div} {}^N \boldsymbol{{\cal T}}^{\sm{(n)}} & := &- \epsilon_{ijk} \Big[ 2 \left(\CD_{2} \bd P^{i \sm{(p)}}\right) \wedge \bd P^{j \sm{(q)}} \wedge \bd X^k \non
   &&\hspace{1cm} + \left(\CD_{3} \bd P^{i \sm{(r)}}\right) \wedge \bd P^{j \sm{(s)}} \wedge \bd P^{k \sm{(t)}} \Big].
   \end{eqnarray}
We have earlier demonstrated the formal equivalence between the Newtonian equations and the relativistic gravitoelectric equations. The generalization of the Newtonian solution scheme to obtain the corresponding relativistic scheme is now straightforward. 

	\subsubsection{Einstein's Theory}
\label{sec:generalsolution}

The perturbative gravitoelectric Lagrange--Einstein system starts at $n=1$ with the Eqs. $\lbrace\eqref{elec1_1}, \eqref{elec2_1}\rbrace$, uniquely determined by the corresponding constraint initial data \eqref{initialdata1}. 
The $n$th--order, $n>1$, gravitoelectric solution scheme reads:
\begin{eqnarray}
    \label{elec1_pertn}  G_{ab}  \bP^{a \sm{(n)}} \wedge \delta^b_{\ j} \bd X^j &=&  \boldsymbol{{\cal S}}^{\sm{(n)}} \;; \\
    \label{elec2_pertn}  \epsilon_{abc}  {\CD_{1}} \bP^{a \sm{(n)}}\wedge \delta^b_{\ j} \bd X^j \wedge \delta^c_{\ k} \bd X^k &=&  \boldsymbol{{\cal T}}^{\sm{(n)}} \;,
\end{eqnarray}
which is uniquely determined by the source terms:
    \begin{eqnarray}
    \boldsymbol{{\cal S}}^{\sm{(n)}} & := & G^{\sm{(r)}}_{ab} \displaystyle\int_{t_0}^t \left( - \dot{\bP}^{a \sm{(s)}} \wedge \bP^{b \sm{(t)}} \right) dt' \;; \\
    \boldsymbol{{\cal T}}^{\sm{(n)}} & := &  -\epsilon_{abc} \Big( 2 \left({\CD_{2}} \mathbf{P}^{a \sm{(p)}}\right) \wedge \mathbf{P}^{b \sm{(q)}} \wedge \delta^c_{\ k} \bd X^k \non
    && \hspace{1cm} + \left({\CD_{3}} \mathbf{P}^{a \sm{(r)}}\right) \wedge \mathbf{P}^{b \sm{(s)}} \wedge \mathbf{P}^{c \sm{(t)}} \Big) \;.
    \end{eqnarray}
The coefficient form of these equations is given by Eqs. \eqref{elec1_pert_coeff_1}--\eqref{eq:trsysGR}.

\section{Application of the solution scheme}
\label{sec:reconstruction}

To illustrate the use of the scheme \system{elec1_pertn}{elec2_pertn} in practice, we will in what follows explicitly explain the construction of relativistic solutions from Newtonian ones for the general procedure and through examples, in Subsections \ref{sec:LPTsol_recovering} and \ref{sec:sol_second}, respectively. Before we do so we explain the general systematics of the solution scheme.

\subsection{Systematics of the solutions}

The $n$th--order scheme is a hierarchy of ordinary second--order differential equations, sourced by an inhomogeneity resulting from combinations of lower--order terms. Thanks to the linearity of the ordinary differential equations (ODEs), the solution is, at any order $n$, a linear superposition of modes that we will label by $l$:
\begin{equation}
\label{sum_modes}
P^{i \sm{(n)}}_{\ j} = \displaystyle\sum \limits_l P^{i \ {\sm{(n,l)}} }_{\ j} \;\;.
\end{equation}
In the Newtonian case, and for the gravitoelectric relativistic part, the modes can be further separated into spatial and temporal parts: $P^{i \ {\sm{(n,l)}} }_{\ j}=\xi^{(n,l)} (t) P^{i \ {\sm{(n,l)}} }_{\ j} (X_{k})$. This is due to the fact that \eqref{elec2_pertn} is an ODE and that its coefficients only depend on time.

From the theory of second--order ODEs it is known (see, e.g., Section 2.1.1 of Ref. \cite{odesols}), that an equation of the form
\begin{equation}
f_{2}\left(a\right){y}^{\prime\prime}+f_{1}\left(a\right){y}^{\prime}+f_{0}\left(a\right)y=g\left(a\right)\;,
\end{equation}
will have as the general solution:
\begin{equation}
\label{gen_sol_ode}
y\left(a\right)=C_{1}y_{1}\left(a\right)+C_{2}y_{2}\left(a\right)+\intop^{a}_{\initial{a}}G\left(a,s\right)g\left(s\right) ds\;,
\end{equation}
where Green's function $G\left(a,s\right)$ is defined by
\begin{equation}
G\left(a,s\right)=\frac{1}{f_{2}\left(s\right)}\;\frac{y_{2}\left(a\right)y_{1}\left(s\right)-y_{1}\left(a\right)y_{2}\left(s\right)}{y_{1}\left(s\right){y}^{\prime}_{2}\left(s\right)-{y}^{\prime}_{1}\left(s\right)y_{2}\left(s\right)}\;.
\label{greensfun}
\end{equation}
Therefore, at any order, the solution will have two modes $l$ that are given by the homogeneous solution, known for a given background model (in the examples we will explicitly give the solutions for the Einstein--de Sitter case, henceforth EdS, and the Cold Dark Matter background with a cosmological constant, henceforth $\Lambda$CDM). The different modes of the particular solution can be calculated from the integral in Eq. \eqref{gen_sol_ode} by setting $g={\cal T}^{\sm{(n)}}$. As integration is linear, the particular solution can be computed for each subpart of the source separately, and those parts appear as a $P^{i \ {\sm{(n,l)}} }_{\ j}$ in the sum \eqref{sum_modes}.

To study these subparts, we split the perturbations into their trace, their symmetric trace--free part and their antisymmetric part:
\begin{equation}
    \bP^a = \frac{1}{3} P \delta^a_{\ j} \bd X^j + \mathbf{\Pi}^{a}+ \boldsymbol{\FP}^{a} \;.
\end{equation}
Then, Eqs. $\lbrace\eqref{eq:asysGR}, \eqref{eq:trsysGR}\rbrace$ read:
\begin{align}
\FP_{i j}^{\sm{(n)}}=\displaystyle\int_{\initial t}^t &\frac{1}{3} \left(G_{a[i}^{\sm{(r)}}\Pi_{\ j]}^{a\sm{(t)}}\dot{P}^{\sm{(s)}}-G_{a[i}^{\sm{(r)}}\dot{\Pi}_{\ j]}^{a\sm{(s)}}P^{\sm{(t)}}\right)\non
+&\frac{1}{3}\left(G_{a[i}^{\sm{(r)}}\FP_{\ j]}^{a\sm{(t)}}\dot{P}^{\sm{(s)}}-G_{a[i}^{\sm{(r)}}\dot{\FP}_{\ j]}^{a\sm{(s)}}P^{\sm{(t)}}\right)\non
+&\left(G_{ab}^{\sm{(r)}}\dot{\FP}_{\ [i}^{a\sm{(s)}}\Pi_{\ j]}^{b\sm{(t)}}-G_{ab}^{\sm{(r)}}\FP_{\ [i}^{b\sm{(t)}}\dot{\Pi}_{\ j]}^{a\sm{(s)}}\right) \non
+&\left(G_{ab}^{\sm{(r)}}\dot{\Pi}_{\ [i}^{a\sm{(s)}}\Pi_{\ j]}^{b\sm{(t)}}+G_{ab}^{\sm{(r)}}\dot{\FP}_{\ [i}^{a\sm{(s)}}\FP_{\ j]}^{b\sm{(t)}}\right) dt' \;\;;
\end{align}
\begin{align}
\CD_{1}&P^{\sm{(n)}} =-\frac{2}{3} P^{\sm{(q)}} \CD_{2}P^{\sm{(p)}} + \Pi_{\ a}^{b\sm{(q)}} \CD_{2}\Pi_{\ b}^{a\sm{(p)}} + \FP_{\ a}^{b\sm{(q)}} \CD_{2}\FP_{\ b}^{a\sm{(p)}}\non
&-\frac{1}{2}\left[\frac{1}{3}\left(\frac{2}{3}P^{\sm{(t)}}P^{\sm{(r)}}-\Pi_{\ b}^{a\sm{(t)}}\Pi_{\ a}^{b\sm{(r)}}-\FP_{\ b}^{a\sm{(t)}}\FP_{\ a}^{b\sm{(r)}}\right)\CD_{3}P^{\sm{(s)}}\right.\non
&+\frac{1}{3}\left(-P^{\sm{(t)}}\FP_{\ b}^{a\sm{(r)}}-\FP_{\ b}^{a\sm{(t)}}P^{\sm{(r)}}\right)\CD_{3}\FP_{\ a}^{b\sm{(s)}}\non
&+\frac{1}{3}\left(-P^{\sm{(t)}}\Pi_{\ b}^{a\sm{(r)}}-\Pi_{\ b}^{a\sm{(t)}}P^{\sm{(r)}}\right)\CD_{3}\Pi_{\ a}^{b\sm{(s)}}\non
&+\left(\Pi_{\ c}^{a\sm{(t)}}\Pi_{\ b}^{c\sm{(r)}}+\Pi_{\ b}^{c\sm{(t)}}\Pi_{\ c}^{a\sm{(r)}}+\Pi_{\ c}^{a\sm{(t)}}\FP_{\ b}^{c\sm{(r)}}
+\Pi_{\ b}^{c\sm{(t)}}\FP_{\ c}^{a\sm{(r)}}\right.\non
&+\FP_{\ c}^{a\sm{(t)}}\Pi_{\ b}^{c\sm{(r)}}\left.+\FP_{\ b}^{c\sm{(t)}}\Pi_{\ c}^{a\sm{(r)}}+\FP_{\ c}^{a\sm{(t)}}\FP_{\ b}^{c\sm{(r)}}+\FP_{\ b}^{c\sm{(t)}}\FP_{\ c}^{a\sm{(r)}}\right)\non
&\left.\left(\CD_{3}\Pi_{\ a}^{b\sm{(s)}}+\CD_{3}\FP_{\ a}^{b\sm{(s)}}\right)\right].
\end{align}
Hence, the trace and the antisymmetric parts are completely determined by the lower--order expressions of all parts (four equations for four components of $P^{i}_{\ j}$). What is missing is an equation for the five components of the trace--free symmetric term $\Pi_{j}^{i\sm{(n)}}$.
Recall that the gravitoelectric system is only closed after imposing the \MRP, which then couples the trace--free symmetric time--evolution to the one of the trace and encodes the spatial dependence in a Poisson equation.

\subsection{Reconstruction of GR solutions}

To illustrate the scheme for the GR case, we will discuss here how to reconstruct the full $n$th--order solution from the recursive equations \system{elec1_pertn}{elec2_pertn}.

\subsubsection{Trace part}
The trace part is the main part that is given by the hierarchy. In the absence of the trace--free symmetric term $\Pi_{j}^{i\sm{(n)}}$, there is no antisymmetric term emerging and we are left with a recursion relation for the trace:
\begin{equation}
\CD_{1}P^{\sm{(n)}}=-\frac{2}{3} P^{\sm{(q)}} \CD_{2}P^{\sm{(p)}} - \frac{1}{9}P^{\sm{(t)}}P^{\sm{(r)}}\CD_{3}P^{\sm{(s)}}\;.
\end{equation}

\subsubsection{Antisymmetric part}

It may appear counterintuitive that a nonvanishing antisymmetric part arises (starting from second order), given our assumption of irrotationality due to the given foliation of spacetime. However, this fact is known from the Newtonian Lagrangian perturbation theory, where antisymmetric parts arise, starting at second order, in Lagrangian space, while no vorticity is created in Eulerian space \cite{buchertehlers}. Our comoving setting corresponds to the Lagrangian picture of fluid motion, and
the antisymmetric terms at order $n$ satisfy and follow from the irrotationality condition \eqref{elec1_pertn}, given all subleading terms $p=1\ldots n-1$.
However, we need to reconstruct a part of the trace--free symmetric term to recover all the Newtonian modes that have antisymmetric components, a problem to which we turn now.

\subsubsection{Trace--free symmetric part}
\label{sec:trfreesym}
As our scheme does not separately provide a relation that determines the five coefficients of the trace--free symmetric part (these equations are part of the gravitomagnetic scheme), we have to reconstruct the relevant part that complies with the Newtonian solutions. To achieve this it suffices to realize that the one--form fields $\bP^a (X^k ,t)$ become integrable in the \MRP, $\bd P^i (X^k,t)$, and so also the trace--free symmetric part. Hence, in the \MRP, the trace--free symmetric part of $\bd P^i (X^k ,t)$ inherits the time--evolution from the trace. With this in mind, and due to the superposition property of our solution scheme, we are entitled to split the general trace--free symmetric coefficients $\Pi_{ij}$ into a part that reproduces the trace--free symmetric part of $\bP^a (X^k ,t)$ in the \MRP, denoted by $^{E}\Pi_{ij}$, and another part $^{H}\Pi_{ij}$. This is possible at any order:
\begin{equation}
\label{sum_modes}
\Pi^{\sm{(n)}}_{ij} = \displaystyle\sum \limits_{l\,m} \xi^{{\sm{(n,l)}}}(t) \,^{E}\Pi^{\ {\sm{(n,l)}} }_{ij}(X^k) +  \, ^{H}\Pi^{\ {\sm{(n,m)}} }_{ij} (X^k, t) .
\end{equation}
The temporal coefficients $\xi^{{\sm{(n,l)}}}$ are the same for the trace and the trace--free symmetric gravitoelectric parts. For the full GR solution there is in addition a contribution,
denoted by $^{H}\Pi_{ij}$, which is related to gravitational waves. We will investigate this part in the followup article.
For the time being we note that the superposition property discussed above assures that the resulting individual terms in the decomposition \eqref{sum_modes} are correct, if we use only this gravitoelectric part of the trace--free symmetric tensor in the hierarchy. Thus, even though the scheme does not determine all the components of $P_{\ j}^{i \sm{(n)}}$ without solving the gravitomagnetic equations, it is consistent for the terms it delivers. 
Moreover, by inspection of corresponding perturbation and solution schemes that we derived for the gravitomagnetic part \cite{alles}, we can conclude that the so--reconstructed solutions provide the leading--order modes of the relativistic solutions at any order.
Of course, inserting the reconstructed solution into the full set of Einstein equations will result in constraints on initial data in addition to the standard constraints. As an example we will discuss the constraints in the first--order scheme given below.

\subsection{Example 1: recovering parts of the general first--order solution}
\label{sec:LPTsol_recovering}

To illustrate the hierarchy we begin with the first--order equations of the scheme \system{elec1_pertn}{elec2_pertn}, i.e. in coefficient form \eqref{elec1_pert_coeff_1}. With the split in space and time coefficients, the latter are the well--known solutions of the equation (equivalent to the equation in the Newtonian scheme \cite{buchert89,bildhaueretal,buchert92}):
\begin{eqnarray}
\ddot{\xi} + 2 H \dot{\xi} -\frac{3}{2} \initial{H}^{2} \initial{\Omega}{}_{m} a^{-3} \xi = W a^{-3}.
\end{eqnarray}
For an EdS universe the modes are proportional to $a$, $a^{-3/2}$ and $a^{0}$. Together with the initial conditions \eqref{initialdata2}, the solution for the trace found from \eqref{elec1_pert_coeff_1} reads:
\begin{align}
P^{\sm{(1)}}=&\frac{3}{5}\left[(U\initial t +\frac{3}{2}W\initial t^2)a\right.\left.-\left(U\initial t -W\initial t^2\right)a^{-\frac{3}{2}}-\frac{5}{2}W\initial t^2\right].
\label{eq:trGRfo}
\end{align}
The antisymmetric part vanishes in view of \eqref{elec1_pert_coeff_1}, $\FP_{\ i}^{a\sm{(1)}}=0$. We then need to reconstruct the trace--free symmetric part along the lines described in \ref{sec:trfreesym} to complete the solution:
\begin{align}
^{E}\Pi_{ij}^{\sm{(1)}}&=\frac{3a}{5}\left(^{E}U^{tl}_{ij} \initial t+\frac{3}{2}{^{E}W}^{tl}_{ij} \initial t^{2}\right)\non
&-\frac{3}{5a^{3/2}}\left(^{E}U^{tl}_{ij} \initial t-{^{E}W}^{tl}_{ij} \initial t^{2}  \right)-\frac{3}{2}{^{E}W}^{tl}_{ij}\initial t^{2}\;.
\label{eq:symtfGRfo}
\end{align}
The notation $^{tl}$ stands for the traceless part. The initial fields have been split accordingly:
\begin{equation}
U_{ij}=:{^{E}U}_{ij}+^{H}U_{ij}\:\:;\:\:W_{ij} = : {^{E}W}_{ij}+^{H}W_{ij}\;,
\label{splitinitial}
\end{equation}
i.e. a part initializing the gravitoelectric, and the gravitomagnetic parts, respectively.

We remark that in Newtonian theory the tidal tensor is written in terms of the gravitational potential $\Phi$: 
\begin{equation}
\label{newtontidal}
- {\cal E}_{ij} =\Phi_{,ij}-\frac{1}{3}\delta_{ij}\nabla^{2}\Phi \;,
\end{equation}
where a comma denotes derivative with respect to Eulerian inertial coordinates.
If we consider the first--order solution (here restricted to the growing mode solution for notational ease),
\begin{equation}
{}^{E}P^{i}_{\ j}{}^{\sm{(1)}}=\frac{3}{2}W^{i}_{\ j}\initial t^{2}\left(a-1\right)\;,
\end{equation}
the first--order gravitoelectric part of the spatially projected Weyl tensor assumes the form (note the conventional sign difference of this geometrical definition with the Newtonian (active) definition of ${\cal E}_{ij}$):
\begin{align}
E^{i}_{\ j}{}^{\sm{(1)}}=&-\ddot{\Pi}^{i}_{\ j}{}^{\sm{(1)}}-2H\dot{\Pi}^{i}_{\ j}{}^{\sm{(1)}}\non
=&-\frac{3}{2}\initial t^{2}\left(\ddot{a}+2H\dot{a}\right)\left(W^{i}_{\ j}-\frac{1}{3}W\delta^{i}_{\ j}\right)\non
=&-\frac{3}{2}\initial t^{2}a\left(\frac{3}{2}\initial H^{2}\frac{1}{a^{3}}\right)\left(W^{i}_{\ j}-\frac{1}{3}W\delta^{i}_{\ j}\right)\non
=&-\frac{1}{a^{2}}\left(W^{i}_{\ j}-\frac{1}{3}W\delta^{i}_{\ j}\right).
\end{align}
We find
\begin{equation}
E_{ij}{}^{\sm{(1)}}=-\left(W_{ij}-\frac{1}{3}W\delta_{ij}\right).
\end{equation}
The trace $W$ does not derive from a potential due to nonintegrability of the field. After executing the \MRP,
we obtain (up to the conventional sign difference), the Newtonian tidal tensor \eqref{newtontidal}.

\noindent
Summarizing: given the formal analogy of the solution schemes discussed in Section~\ref{sec:gesolscheme},
the above solution solves the gravitoelectric part of the corresponding
relativistic equations (\ref{elec1_pert_coeff_1}).
The trace--free symmetric part \eqref{eq:symtfGRfo}, however,  is only a part of the solution in the relativistic case.
Equation (102) in Ref. \cite{rza1} states that the first--order equation for the relativistic trace--free symmetric part reads:
\begin{eqnarray}
\ddot{\Pi}_{ij}^{\sm{(1)}}+3H\dot{\Pi}_{ij}^{\sm{(1)}}-a^{-2}\Pi_{ij\ |k}^{|k\sm{(1)}}\non
=-a^{-2}\left(\ST_{ij}+P_{|ij}^{\sm{(1)}}-\frac{1}{3}P_{\ |k}^{|k\sm{(1)}}\delta_{ij}\right)\;,
\label{eq:tf_sym_comp}
\end{eqnarray}
where $\ST_{ij}$ is the trace--free part of the initial Ricci tensor.
Plugging Eqs. \eqref{eq:trGRfo} and \eqref{eq:symtfGRfo} into Eq. \eqref{eq:tf_sym_comp}, we can check whether our relativistic generalization satisfies the full equation. Three modes appear in the equation:
$a^{-2}$, $a^{-1}$ and $a^{-7/2}$. The equation has to be satisfied
at any time, and thus each mode must lead to cancellation of the coefficients. This leads to the
following constraints ($\initial H := 2/3\initial t$): 
\begin{eqnarray}
^{E}\ST_{ij}=-\initial H ^{E}U^{tl}_{(ij)} - ^{E}W^{tl}_{(ij)} \;\;;\qquad\qquad\non
^{E}U_{(ij)\ |k}^{|k}={^{E}U}_{\ k|ij}^{k}\;\;\;;\;\;\;
^{E}W_{(ij)\ |k}^{|k}={^{E}W}_{\ k|ij}^{k}\;\;.\qquad\qquad
\label{N_to_GR_constr}
\end{eqnarray}
The first equation corresponds to the definition of the trace--free part of the initial Ricci tensor, Eq.~\eqref{kinematicRicci} in the EdS case studied here.
In view of the constraints $U_{[ij]} = 0$ and $W_{[ij]} = 0$ ({\it cf.} \eqref{initialdata1}), the other two conditions are equivalent to:
\begin{equation}
^{E}U_{ij\ |k}^{|k}={^{E}U}_{\ k|ij}^{k}\;\;\;;\;\;\;
^{E}W_{ij\ |k}^{|k}={^{E}W}_{\ k|ij}^{k}\;\;.
\label{MMM}
\end{equation}
What we call gravitoelectric part in the decomposition of initial conditions \eqref{splitinitial} is therefore determined to be the one that solves \eqref{N_to_GR_constr}. The part contributing to the propagating gravitomagnetic part is then its complement. This labelling is not completely unambiguous, because in this scheme, the gravitomagnetic part computed from the gravitoelectric part is not null, see below for the first--order scheme. (Nevertheless, as we will show in the followup paper, it generates a null dynamical Ricci curvature tensor.)

To check how constraining these relations are, beyond the constraints that we already have, we consider the first and second time--derivatives of the momentum constraints and evaluate them at initial time in order to obtain constraints on the initial fields. Taking the second spatial derivative of these equations and contracting them with respect to one index, we get for $U_{ij}$:
\begin{equation}
U_{\ j|k}^{k}=U_{\ k|j}^{k}\Rightarrow U_{\ j|ik}^{k}=U_{\ k|ij}^{k}\Rightarrow U_{\ j|ik}^{k|i}=U_{\ k|ij}^{k|i}\;.
\end{equation}
The latter identity is solved by the gravitoelectric and the gravitomagnetic parts independently. For the gravitoelectric part, we have:
\begin{equation}
^{E}U_{\ j|ik}^{k|i}={^{E}U}_{\ k|ij}^{k|i}\;,
\end{equation}
which is equal to the once contracted spatial derivative of the above constraint \eqref{MMM}. We conclude that Eq. \eqref{MMM} and the momentum constraints are compatible with but not equivalent to our constraints. They have to be solved independently in order for the solution to be compatible with both the evolution equation and the momentum constraints.

What they do constrain are derivatives of the gravitomagnetic part. To derive these constraints, let us first note that the first--order expression for the magnetic part can be found from Eq. (107) in Ref. \cite{rza1}:
\begin{equation}
{}^E {H}_{ij} = a(t) \epsilon^{sl}_{\ \ (i}{}^E {\dot{\Pi}}_{j)l\vert s}\:\:.
\label{HP}
\end{equation}
The solution for ${}^E \Pi_{ij}$,  {\it cf.}~\eqref{eq:symtfGRfo}, shows that spatial derivatives of the first--order magnetic part can be traced back to spatial derivatives of $U_{ij}$ and $W_{ij}$. Together with the first--order momentum constraints, $\dot{P}_{\ i|j}^{i}=\dot{P}_{\ j|i}^{i}$, and imposing the constraints \eqref{N_to_GR_constr}, we get:
\begin{equation}
\epsilon^{uri}\ \epsilon^{sl}_{\ \ (i} {}^E {U}_{j)s \vert l r} = 0\;\;;\;\;\epsilon^{uri}\ \epsilon^{sl}_{\ \ (i} {}^E {W}_{j)s \vert l r} = 0\:\:.
\label{UWconstraint}
\end{equation}
Thus, via \eqref{eq:symtfGRfo}, this leads to
\begin{equation}
\epsilon^{uri}\  {}^E {H}_{ij \vert r} = 0\:\:,
\label{idH}
\end{equation}
i.e., the curl of ${}^E {H}_{ij}$ vanishes.

For its divergence the constraints \eqref{N_to_GR_constr} are not necessary. Taking the divergence of \eqref{HP}, and using the momentum constraints in the form  ${}^{E}\dot{\Pi}_{\ l\vert is}^{i}=2/3 \dot{P}_{\vert ls}$, we can show:
\begin{equation}
^{E} {H}_{ij}{}^{\vert i} = 0\:\:.
\label{idH3}
\end{equation}
By combining \eqref{idH} and \eqref{idH3}, we conclude that  
\begin{equation}
\Delta_0 {}^{E} {H}_{ij}= 0\:\:.
\label{idH4}
\end{equation}
Thus, the gravitomagnetic part that is generated by the gravitoelectric part is a harmonic tensor field at first order.
This harmonic field can be constrained in the initial conditions (removed) by topological conditions on the perturbations. In an upcoming article, we will discuss such conditions. 
 
\subsection{Example 2: constructing second--order solutions for `slaved initial data'}
\label{sec:sol_second}

Let us now write out the system \system{elec1_pertn}{elec2_pertn} explicitly for $n=2$. We simplify the first--order source by imposing the so--called `slaving condition' $U^{i}_{\ j}=W^{i}_{\ j}\initial t$ (as explained in Refs. \cite{buchert89,buchert92}, for second order in Ref. \cite{buchert93L}, and used to present the third-- \cite{buchert94} and fourth--order \cite{rampfbuchert} Newtonian solutions). This is not necessary but increases readability. The sum of \eqref{eq:trGRfo} and \eqref{eq:symtfGRfo} becomes:
\begin{equation}
P_{ij}^{\sm{(1)}}=\frac{3}{2}W_{ij}\initial t^{2}\left(a-1\right)\;.
\end{equation}
At second order \eqref{eq:trsysGR} is simply
\begin{equation}
{\CD_{1}} P^{i \sm{(2)}}_{\ i} =- \left( {\CD_{2}} P^{i \sm{(1)}}_{\ i} \right) P^{j \sm{(1)}}_{\ j}  + \left( {\CD_{2}} P^{i \sm{(1)}}_{\ j} \right) P^{j \sm{(1)}}_{\ i} \;,
\end{equation}
and we have the system:
\begin{eqnarray}
\left\{ \begin{split} & \ddot{\xi}^{\sm{(2)}}+2\frac{\dot{a}}{a}\dot{\xi}^{\sm{(2)}}+3\frac{\ddot{a}}{a}\xi^{\sm{(2)}}=\frac{3}{4}\initial t^{2}\left(a^{-1}-a^{-3}\right)\;;\\
 & C^{\sm{(2)}}=W_{\  j}^{i}W_{\  i}^{j}-WW\;,
\end{split}	
\right.
\end{eqnarray}
with the source $g^{\sm{(2)}}(t)=\frac{3}{4}\initial t^{2}\left(a^{-1}-a^{-3}\right)$.

To systematically determine the temporal coefficients of the hierarchy, it is useful to
write the operator $\CD_{1}$ in terms of $a$. We find:
\begin{eqnarray}
&&g\left(a\right) = \initial{\Omega}_{m}\initial H^{2} \times \nonumber \\
&&\left(\left(\frac{1}{a} +a^{2}c\right)P^{\prime\prime}(a)+\frac{3}{2}\left(\frac{1}{a^{2}}+2ac\right)P^{\prime}(a)-\frac{3}{2a^{3}}P(a)\right),\nonumber\\
\end{eqnarray}
where $c=\initial{\Omega}_{\Lambda}/\initial{\Omega}_{m}$. For an EdS background, $c=0$, the homogeneous solution is
\begin{equation}
D\left(a\right)=aC_{1}+a^{-3/2}C_{2}\;;
\label{growthEdS}
\end{equation}
Green's function of Eq. \eqref{greensfun} is
\begin{equation}
G\left(a,s\right)=\frac{2}{5}\frac{s\left(a^{5/2}-s^{5/2}\right)}{\initial{\Omega}_{m}\initial H^{2}a^{3/2}}\;.
\label{greenEdS}
\end{equation}
Now, it is a matter of a simple integration and
Eq.~\eqref{gen_sol_ode} gives the second--order trace solution:
\begin{equation}
P^{\sm{(2)}}={}^{1}C^{\sm{(2)}}a+{}^{2}C^{\sm{(2)}}a^{-3/2}+\frac{9}{8}\initial t^{4}\left(1+\frac{3}{7}a^{2}\right)C^{\sm{(2)}}\;.
\end{equation}
To find the spatial coefficients of the solution, we use the initial values for the coframe and its
time--derivative. They have been chosen to vanish for all orders higher than one in the hierarchy of solutions
of Eqs.~(\ref{eq:trsysGR}). Therefore, we find the system:
\begin{eqnarray}
P^{\sm{(2)}}(\initial t)={}^{1}C+{}^{2}C+\frac{45}{28}C^{\sm{(2)}}\initial t^{4}=0\;;\non
\dot{P}^{\sm{(2)}}(\initial t)=\frac{2}{3\initial t}{}^{1}C-\frac{1}{\initial t}{}^{2}C+\frac{9}{14}C^{(2)}\initial t^{4}=0\;,
\end{eqnarray}
which fixes all constants to be $\propto C^{\sm{(2)}}$.
Thus, the second--order trace solution reads:
\begin{eqnarray}
P^{\sm{(2)}}&=&\xi^{\sm{(2)}}_{+} \left(WW - W_{\ j}^{i}W_{\ i}^{j}\right)\;; \nonumber \\
\xi^{\sm{(2)}}_{+}&=&\frac{9}{4}\initial t^{4}\left(-\frac{3}{14}a^{2}+\frac{3}{5}a-\frac{1}{2}+\frac{4}{35}a^{-3/2}\right).
\end{eqnarray}
After executing the \MR this coincides with the second--order Newtonian solution of Ref. \cite{buchertehlers}.

The antisymmetric equation \eqref{elec1_pertn} still delivers 
$\FP_{\ i}^{a\sm{(2)}}=0$. This is due to the restriction to `slaved initial conditions', otherwise we would have a nonvanishing part here.
Thus, we only need the trace--free symmetric part to complete the solution. The gravitoelectric part can be written as
\begin{equation}
^{E}\Pi_{ij}^{\sm{(2)}}=\xi^{\sm{(2)}}_{+} \CS^{(2)}_{ij}\;,
\label{secondslave}
\end{equation}
where the trace of $\CS^{(2)}_{ij}$ is given by $\left(W^2 - W_{ \ j}^{i}W_{ \ i}^{j}\right)\initial t^2$. The rest of its components can be determined from
the generalization $\CS^{(2)}_{|ij}\rightarrow\CS^{(2)}_{ij}$, where $\CS^{(2)}$ is the solution to the Newtonian Poisson equation $\Delta_0 \CS^{(2)}=\left((W^k_{\ | k})^2 - W_{\ | j}^{i}W_{ \ | i}^{j} \right)\initial t^2$, and where $\Delta_0$ denotes the Laplacian in local (Lagrangian) coordinates (see Ref. \cite{buchertehlers}). To avoid passing by the generalization of the Newtonian result, one can of course also insert $^{E}\Pi_{ij}^{\sm{(2)}}$ into \eqref{secondslave} and solve the remaining relativistic equations of the gravitomagnetic part to find the off--trace components of $\CS^{(2)}_{ij}$. The gravitomagnetic contribution encoded in $\CS^{(2)}_{ij}$ has recently been studied to second order in Ref. \cite{rampfwiegand}.

The explicit derivation of the inhomogeneous second--order term in this subsection illustrates that, using \eqref{gen_sol_ode} and \eqref{greenEdS}, the calculation of the temporal evolution of the general relativistic trace part is straightforward and only involves the calculation of integrals.
This can also be easily extended to perturbations of a $\Lambda$CDM universe model by noting that \eqref{growthEdS} becomes:
\begin{equation}
D\left(a\right)=a\,^{(2)}F_{1}\left(\frac{1}{3},1,\frac{11}{6};-ca^{3}\right)C_{1}+\sqrt{\frac{1}{a^{3}}+c}\,C_{2}\;,
\label{growthLCDM}
\end{equation}
with the Gauss hypergeometric function $^{(2)}F_{1}$. Greens' function reads in this case:
{\small
\begin{equation}
G\left(s,a\right)=\frac{2}{5}\frac{s}{\initial{\Omega}_{m}\initial H^{2}}\left(D_{+}\left(a,c\right)-D_{+}\left(s,c\right)\sqrt{\frac{\left(1+ca^{3}\right)s^{3}}{\left(1+cs^{3}\right)a^{3}}}\right),
\end{equation}}
where $D_{+}\left(a,c\right)$ is the first term in \eqref{growthLCDM}.

\section{Summary and concluding remarks}
\label{sec:conclusion}

We have investigated gravitoelectric perturbation and solution schemes at any order in relativistic Lagrangian perturbation theory. These schemes cover the full hierarchy of the Newtonian Lagrangian perturbation theory if restricted to integrable Cartan coframe fields. 

Despite the fact that the solution scheme presented in this work gives on its own not all parts of the relativistic perturbation solutions, it delivers an important part relevant to the formation of large--scale structure. As is well--known (see, e.g., discussions in Refs. \cite{lischwarz} and \cite{alles}), the fastest growing scalar modes of the GR solutions correspond to the Newtonian modes, shown up to second order and, by inspection of the schemes we investigated, we showed
this to hold for the gravitoelectric part also beyond second order. As we recover all the Newtonian terms with their correct temporal evolution and their constrained spatial coefficients, we also know that our solution contains all terms that become important in the Late Universe. The presented scheme is explicit enough to derive solutions at any desired order by algebraic codes along the lines of the reconstruction rules that we exemplified up to the second order.

We demonstrated the close formal correspondence of the gravitoelectric Lagrange--Einstein system to the Newtonian theory furnishing construction rules that also allow to find other, nonperturbative relativistic solutions from Newtonian ones. 

The role of gravitational waves, corresponding to the missing part in our scheme, has to be further explored. 
The missing part, which we denoted by $^{H}\Pi_{ij}$ in the coefficients of the trace--free symmetric parts of the perturbations, corresponds at first order to `free gravitational waves', i.e. that part of gravitational radiation that does not scatter at the sources. This changes at higher orders, since this part will couple to the sources starting at second order. We will investigate in detail the general first--order scheme including gravitational waves in the next article of this series, where we also identify the transformations and restrictions that have to be imposed to obtain the known solutions of the standard perturbation theory, where perturbations are embedded into the background spacetime. 

\vspace{-12pt}

\subsection*{Acknowledgements}

\vspace{-6pt}

{\small The work of AA, TB and FR was conducted within the ``Lyon Institute of Origins'' under grant ANR--10--LABX--66.
AA and FR acknowledge support by the {\'E}cole Doctorale Lyon. TB and FR acknowledge student exchange support from the French--Bavarian Cooperation Center, BFHZ Munich \url{http://www.bayern-france.org/}. AW acknowledges the hospitality of the Harvard CfA where part of this work was done.
Part of the work of AW was supported by the German research organization DFG, Grant--No. WI 4501/1--1.\\
Special thanks go to L\'eo Brunswic, Martin Kerscher, Miko\l aj Korzy\'nski, Pierre Mourier, Slava Mukhanov, Jan Ostrowski, Cornelius Rampf, Herbert Wagner and the referee for fruitful discussions and valuable remarks.  
}{\small \par}

\vspace{-8pt}

\appendix

\section{Equivalence of integrability of the coframes and the flatness of space}
\label{sec:appA}

\vspace{-4pt}

The standard choice of {\it orthonormal coframes} ${\tilde\bet}^a$in the Cartan formalism implies for the spatial metric coefficients ${\tilde g}_{ij} = \delta_{ab} \tilde{\eta}^a_{\ i} \tilde{\eta}^b_{\ j}$, with $\tilde{\eta}^a_{\ i}(t_i) \ne \delta^a_{\ i}$ at initial time, in order to have an initially nontrivial metric.

The alternative choice of {\it adapted coframes} $\bet^a$, used in this article, represents the metric coefficients as $g_{ij} = G_{ab} \eta^a_{\ i} \eta^b_{\ j} $,
where we are entitled to require $\eta^a_{\ i}(t_i) = \delta^a_{\ i}$ at initial time, encoding the initial metric into the coefficients $G_{ab}$, i.e. $G_{ij} = G_{ab}\delta^a_{\ i}\delta^b_{\ j}$. This makes the comparison with the Newtonian choice of Lagrangian coordinates to coincide with the Eulerian ones at some initial time more direct.

As we discussed in \eqref{apA}, the basic assumption is that both coframe types describe the same metric form, i.e. 
$\mathbf{g} = \delta_{cd} \tilde{\boldsymbol{\eta}}^c \otimes \tilde{\boldsymbol{\eta}}^d = G_{ab}  {\boldsymbol{\eta}}^a \otimes {\boldsymbol{\eta}}^b$, from which we infer:
\begin{equation}
G_{ab} = \delta_{cd} \ \tilde{{\eta}}^c_{\ a}\tilde{{\eta}}^d_{\ b} \:\:,
\label{plorki}
\end{equation}
where we denote with $\tilde{{\eta}}^c_{\ a} = e^{\;i}_{c}\tilde{{\eta}}^c_{\; i}$
the coefficients of the projection of $\tilde{\boldsymbol{\eta}}^c$ onto the basis ${\boldsymbol{\eta}}^a$. 

The \MR applied to either of these coframes requires them to be exact forms, ${\tilde\bet}^a = {\bd} {\tilde f}^a$ or ${\bet}^a = {\bd} {f}^a$. 
They then define some global Eulerian coordinates, $\tilde{x}^a$ and  ${x}^a$, respectively.
In the \MRP, Eq. \eqref{plorki} is equivalent to 
\begin{equation}
G_{ab} = \delta_{cd} \ \frac{\partial \tilde{f}^c}{\partial {x}^a}\ \frac{\partial \tilde{f}^d}{\partial {x}^b} \:\:.
\label{plorki2}
\end{equation}
We infer from \eqref{plorki2} that the coefficients $G_{ab}$ just depend on initial vector displacements after executing the \MRP. They are related to the initial deformation gradient in the orthonormal description, as can be seen by looking at the metric equivalence relation in an exact Lagrangian basis,
${\bg} = {\tilde g}_{ij} \bd {\tilde X}^i \otimes \bd {\tilde X}^j = g_{ij} \bd X^i \otimes \bd X^j$: 
\begin{eqnarray}
\label{a1}
{\bg}(X^k , t_i) = \delta_{ab} \tilde{f}^a_{\ | {\tilde X}^i} ({\tilde X}^k , t_i) \tilde{f}^b_{\ | {\tilde X}^j} ({\tilde X}^k , t_i) \bd {\tilde X}^i \otimes \bd {\tilde X}^j 
\non =\delta_{ab} \tilde{f}^a_{\ | i} ({X}^k , t_i) \tilde{f}^b_{\ | j} ({X}^k , t_i) \bd {X}^i \otimes \bd {X}^j 
\non = G_{ab} (X^k ){\delta}^a_{\ i}  {\delta}^b_{\ j} \bd X^i \otimes \bd X^j \;,\qquad
\end{eqnarray}
where a slash denotes derivative with respect to the coordinates $X^i$, as in the main text, and it is explicitly noted otherwise.
From \eqref{plorki2} we conclude that
\begin{equation}
\mathbf{g} =G_{ab}  {\boldsymbol{\eta}}^a \otimes {\boldsymbol{\eta}}^b =  \delta_{cd} \frac{\partial \tilde{f}^c}{\partial {x}^a}\frac{\partial \tilde{f}^d}{\partial {x}^b}  {\bd} {x}^a \otimes  {\bd} {x}^b = \delta_{cd}  {\bd} {\tilde x}^c  \otimes {\bd} {\tilde x}^d \,,
\end{equation}
which is the Euclidean metric. 

Summarizing: execution of the \MR leads, in either of the chosen coframes, to a metric that is equivalent to the Euclidean metric. 
The coefficients $G_{ab}$ can then be expressed in terms of initial vector displacements, {\it cf.} Eq. \eqref{a1}.

\vspace{-15pt}

\section{Erratum and Remarks}
\label{sec:appB}

\vspace{-4pt}

We correct a mistake in the paper \cite{ehlersbuchert}, add a clarification to the paper \cite{rza1} concerning the gravitomagnetic part of the spatially projected Weyl tensor.

\vspace{-6pt}

\subsection{Newtonian perturbation scheme, Ref. \cite{ehlersbuchert}}

The perturbative system of equations derived from Lagrangian Newtonian theory leads to $\lbrace$\eqref{curl_lagr_form_pert},\eqref{div_lagr_form_pert}$\rbrace$:
\begin{eqnarray}
	& \delta_{ij} & \bd \dot{P}^i \wedge \left( \bd X^j +  \bd P^j \right) = \delta_{ij} a^{-2} \bd U^i \wedge \bd X^j \;; \\
	& \epsilon_{ijk} & \Bigg[ \left( \left( {D} - 4 \pi G \varrho_H \right) \bd P^i \right) \wedge \bd X^j \wedge \bd X^k \non
	& + & \left( \left( 2 {D} - 4 \pi G \varrho_H \right) \bd P^i \right) \wedge \bd P^j \wedge \bd X^k \non
	& + & \left( \left( {D} - \frac{4 \pi G}{3} \varrho_H \right) \bd P^i \right) \wedge \bd P^j \wedge \bd P^k \Bigg] \non
	& = & - \epsilon_{ijk} \frac{4 \pi G}{3} \delta \initial \varrho a^{-3} \bd X^i \wedge \bd X^j \wedge \bd X^k \;,
\end{eqnarray}
where the operator ${D} := d^2 / dt^2 + 2 H d / dt$.
The numerical coefficients of this system differ from the result presented in Ref. \cite{ehlersbuchert}: Eqs. (28a--d).

\subsection{Gravitomagnetic part of the spatially projected Weyl tensor, Ref. \cite{rza1}}

\vspace{-3pt}

In Ref. \cite{rza1} the symmetrization of the spatial parts of the Weyl tensor has not always been written explicitly, which may lead to confusion.
The idea of a not manifestly symmetric writing is best seen in the equations for the gravitoelectric part: its definition in Eq. \eqref{ge} already assumes that the field equations hold, which are then recovered by the conditions \eqref{ge_sys}. While having advantages, this representation is implicit. 
The same applies for the gravitomagnetic part (see Ref. \cite{rza1}: Eqs. (70,73), where the momentum constraints have been inserted and recovered through an explicit symmetry condition). If we wish to consider the original geometrical definition of these tensors without inserting the field equations, then we have to write, e.g. for the gravitomagnetic part (here written for the adapted coframes used in this article):
\begin{align}
H_{ij} =& - \frac{1}{J} \epsilon^{nkl} g_{n(i} \Theta_{ \textbf{(}j)k \textbf{)} ||l} \non
=& - \frac{1}{2J} \epsilon^{nkl} g_{n(i} G_{ab} \left( \left( \eta^a_{\ j)} \dot{\eta}^b_{\ k} \right)_{||l} + \left( \eta^a_{\ k} \dot{\eta}^b_{\ j)} \right)_{||l} \right) \non
=& - \frac{1}{2J} \epsilon^{nkl} g_{n(i} G_{ab} \left( \eta^a_{\ j)||l} \dot{\eta}^b_{\ k} + \eta^a_{\ j)} \dot{\eta}^b_{\ k||l} \right. \non
& \hspace{2.5cm} \left.+ \eta^a_{\ k||l} \dot{\eta}^b_{\ j)} + \eta^a_{\ k} \dot{\eta}^b_{\ j)||l} \right) \;,
\end{align}
where $g_{ni} = G_{ab}\eta^a_{\ n}\eta^b_{\ i}$.
Employing the irrotationality condition \eqref{irrotationality2} and taking its covariant derivative,
\begin{align}
G_{ab}\,\dot{\eta}_{\ [i}^{a}\eta_{\ j]||k}^{b}+G_{ab}\,\eta_{\ [j}^{a}\dot{\eta}_{\ i]||k}^{b}=0\;,
\end{align}
we obtain an expression for $H_{ij}$ that manifestly vanishes with the passage to the Newtonian theory through the \MRP: the covariant derivatives of the (now integrable) coframes vanish:
\begin{align}
H_{ij} =-\frac{1}{J}\epsilon^{nkl}g_{n(i}G_{ab}\left(\eta_{\ j)}^{a}\dot{\eta}_{\ k||l}^{b}\right)\; \quad \rightarrow \quad H^N_{j} = \qquad\non
- \frac{1}{{}^N \! J} \epsilon^{nkl} G_{de} G_{ab} f^d_{\ |n} f^e_{\ |(i}f^a_{\ |j)} \left( \dot{f}^b_{\ |kl} - \dot{f}^b_{\ |m} h_{,c}^{\ m} f^c_{\ |kl} \right) =0\;,
\end{align}
which is the result claimed in Ref. \cite{rza1}: Eq. (84), and which is expected in a concise execution of the Newtonian limit \cite{ehlersbuchert_weyl}.

\vspace{-10pt}

\subsection{Remark on the Newton--GR ``dictionary''}

\vspace{-4pt}

The reader may have noticed that the Newtonian tensors which correspond to the GR tensors in the \MR display mixed indices, e.g. in Subsection~\ref{MRexecute} for the \MR of the gravitoelectric part $E^i_{\ j}$. We explain here why this is the case.

Consider, e.g., the Newtonian field strength gradient, $g^a_{\ ,b}$, where a comma denotes a derivative with respect to Eulerian coordinates. We have on purpose denoted the vector index by a counter index here, counting the number of vector components, and we have also used the counter index for the Eulerian derivative, since both become noncoordinate indices by executing the inverse \MRP, i.e., by passing from the Newtonian deformation gradient with respect to Lagrangian coordinates, $f^a_{\ |k}$, to the nonintegrable coframe coefficients, $\eta^a_{\ k}$. This expresses the fact that the vector embedding space disappears through this operation; in the integrable case, the components $f^a$ can be considered as coordinate functions with the coordinates $x^{a \rightarrow i}$. Now, the transformation of the field strength gradient to Lagrangian coordinates involves the inverse of the transformation $f^a$, which we denote by $X^i = h^i (x^b ,t)$: $g^a_{\ ,b} = {\ddot f}^a_{\ | i} h^i_{\ ,b}$, where we have also inserted the definition $g^a = {\ddot f}^a$.
Denoting the coefficients of the inverse of the coframes by $e_b^{\ i}$, we see that the inverse \MR constructs the relativistic analog to $g^a_{\ ,b}$, which is ${\cal F}^a_{\ b} :={\ddot\eta}^a_{\ i} e_b^{\ i}$. Projecting this latter field onto the exact basis with the help of the coframes and their inverse (the frames), we obtain a field with mixed indices, 
${\cal F}^a_{\ b} e_a^{\ i}\eta^b_{\ j} = {\cal F}^i_{\ j}$. The tensor coefficients are then obtained by lowering an index with the spatial metric, ${\cal F}_{kj} = g_{ki}{\cal F}^i_{\ j} =
G_{ab}\eta^a_{\ k} {\ddot \eta}^b_{\ j}$. For mixed indices, we obtain (compare Eq.~\eqref{GRgrav}): 
\begin{equation}
\CF^i_{\ j} = e_a^{\ i} {\ddot\eta}^a_{\ j} =  \frac{1}{2J} \epsilon_{abc} \epsilon^{ikl} {\ddot{\eta}}^{a}_{\ j} {\eta}^{b}_{\ k} {\eta}^{c}_{\ l} \;\;.
\end{equation}




\begin{thebibliography}{2015}

\bibitem[L1]{rza1} T. Buchert and M. Ostermann: Lagrangian theory of structure formation in relativistic cosmology I: Lagrangian framework and definition of a nonperturbative approximation. {\sl Phys. Rev. D} \textbf{86}, 023520 (2012).

\bibitem[L2]{rza2} T. Buchert, C. Nayet, and A. Wiegand: Lagrangian theory of structure formation in relativistic cosmology II: average properties of a generic evolution model. 
{\sl Phys. Rev. D} \textbf{87}, 123503 (2013).

\bibitem{alles} A. Alles: Inhomogeneous Cosmology: an answer to the Dark Matter and Dark Energy problems? {\sl Ph.D. thesis}, Universit\'e Lyon 1 (2014).

\bibitem{Bardeen80} J.M. Bardeen: Gauge--invariant cosmological perturbations. 
{\sl Phys. Rev. D} \textbf{22}, 1882 (1980).

\bibitem{barnesrowlingson} 
A. Barnes and R.R. Rowlingson: Irrotational perfect fluids with a purely electric Weyl tensor.
{\sl Class. Quant. Grav.} \textbf{6}, 949 (1989).

\bibitem{bertschingerhamilton}
E. Bertschinger and A.J.S. Hamilton: Lagrangian evolution of the Weyl tensor.
{\sl The Astrophys. J.} \textbf{435}, 1 (1994).

\bibitem{bildhaueretal}
S. Bildhauer, T. Buchert and M. Kasai: Solutions in Newtonian cosmology -- the pancake theory with cosmological constant.  
{\sl Astron  Astrophys.} \textbf{263},  23  (1992).

\bibitem{bruni:silent} M. Bruni, S. Matarrese and O. Pantano: Dynamics of silent universes. 
{\sl Astrophys. J.} \textbf{445}, 958 (1995).

\bibitem{buchert89} T. Buchert:  A class of solutions in Newtonian cosmology and the pancake theory.
{\sl Astron. Astrophys.} \textbf{223}, 9 (1989). 

\bibitem{buchert92} 
T. Buchert: Lagrangian theory of gravitational instability of Friedmann--Lema\^\i tre cosmologies and the `Zel'dovich approximation'.
{\sl Mon. Not. Roy. Astr. Soc.} \textbf{254}, 729 (1992). 

\bibitem{buchert93L}
T. Buchert: Lagrangian perturbation theory -- a key--model for large--scale structure.
{\sl Astron  Astrophys.} \textbf{267}, L51 (1993). 

\bibitem{buchert94} 
T. Buchert: Lagrangian theory of gravitational instability of Friedmann-Lema\^\i tre cosmologies: generic third--order model
for nonlinear clustering. 
{\sl Mon. Not. Roy. Astron. Soc.} \textbf{267}, 811 (1994).

\bibitem{buchert:lagrangian}
T. Buchert: Lagrangian Perturbation Approach to the Formation of Large--scale Structure. In {\sl Proceedings of the international school of physics Enrico
  Fermi. Course CXXXII: Dark matter in the {U}niverse}, edited by S. Bonometto,
  J. Primack and A. Provenzale, {\sl IOP Press, Amsterdam}  (Societ{\`a} Italiana di Fisica, Varenna sul Lago di Como, 1996).

\bibitem{buchert:focus} 
T. Buchert: Toward physical cosmology: focus on inhomogeneous geometry and its nonperturbative effects. 
{\sl Class. Quant. Grav.} \textbf{28}, 164007 (2011).

\bibitem{buchertehlers} 
T. Buchert and J. Ehlers: Lagrangian theory of gravitational instability of Friedmann--Lema\^\i tre cosmologies -- second order approach: an improved model for nonlinear clustering.
{\sl Mon. Not. Roy. Astron. Soc.} \textbf{264}, 375 (1993).

\bibitem{buchertehlers:average} 
T. Buchert and J. Ehlers: Averaging inhomogeneous Newtonian cosmologies.
{\sl Astron. Astrophys.} \textbf{320}, 1 (1997). 

\bibitem{buchertgoetz}
T. Buchert and G. G\"otz: A class of solutions for self--gravitating dust in Newtonian gravity. 
{\sl J. Math. Phys.} \textbf{28}, 2714 (1987).

\bibitem{chandra}
S. Chandrasekhar : The Mathematical Theory of Black Holes.
{\sl Clarendon Press, Oxford} (1998).

\bibitem{Durrer} R. Durrer: Gauge invariant cosmological perturbation theory.
{\sl Fund. Cosmic Phys.} \textbf{15}, 209 (1994).
 
\bibitem{ehlersbuchert}
J. Ehlers and T. Buchert:
Newtonian cosmology in Lagrangian formulation: foundations and perturbation theory.
{\sl Gen. Rel. Grav.} \textbf{29}, 733 (1997).

\bibitem{ehlersbuchert_weyl}
J. Ehlers and T. Buchert: On the Newtonian limit of the Weyl tensor.
{\sl Gen. Rel. Grav.} \textbf{41}, 2153 (2009).

\bibitem{ellisbruni1}
G.F.R. Ellis, M. Bruni: Covariant and gauge--invariant approach to cosmological density fluctuations. 
{\sl Phys. Rev. D} \textbf{40}, 1804 (1989).

\bibitem{ellisbruni2}
G.F.R. Ellis, J. Hwang, M. Bruni: Covariant and gauge--independent perfect fluid Robertson--Walker perturbations. 
{\sl Phys. Rev. D} \textbf{40}, 1819 (1989). 

\bibitem{henk:silent}
H. van Elst, C. Uggla, W.M. Lesame, G.F.R. Ellis and R. Maartens:
Integrability of irrotational silent cosmological models.
{\sl Class. Quant. Grav.} \textbf{14}, 1151 (1997).

\bibitem{gourgoulhon07}
E. Gourgoulhon: 3+1 Formalism and Bases of Numerical Relativity. Lecture Notes: {\sl ArXiv: gr--qc/0703035} (2007).

\bibitem{gron}
\O. Gr\o n, and H. Sigbjorn : Einstein's General Theory of Relativity.
{\sl Springer, Berlin, Hamburg} (2007).

\bibitem{Kasai95} M. Kasai: Tetrad--based perturbative approach to inhomogeneous universes: A general relativistic version of the Zel'dovich approximation. {\sl Phys. Rev. D} \textbf{52}, 5605 (1995). 

\bibitem{Kodama84} H. Kodama and M. Sasaki: Cosmological Perturbation Theory.{\sl Progress of Theoretical Physics Supplement} \textbf{78}, 1 (1984). 

\bibitem{kofmanpogosyan} L. Kofman and D. Pogosyan: Dynamics of gravitational instability is nonlocal.
{\sl Astrophys. J.} \textbf{442}, 30 (1995).

\bibitem{kopeikinpetrov}
S.M. Kopeikin, A.N. Petrov: Dynamic field theory and equations of motion in cosmology.
{\sl Ann. of Phys.} \textbf{350}, 379 (2014).

\bibitem{lischwarz}
N. Li and D.J. Schwarz: On the onset of cosmological backreaction.
{\sl Phys. Rev. D} \textbf{76}, 083011 (2007).

\bibitem{Matarrese94a} S. Matarrese, O. Pantano and D. Saez: General relativistic dynamics of irrotational dust: Cosmological implications. {\sl Physical Review Letters} \textbf{72}, 320--323 (1994).

\bibitem{Matarrese94b} S. Matarrese, O. Pantano, and D. Saez: A relativistic approach to gravitational instability in the expanding Universe -- Second--order Lagrangian solutions. {\sl Monthly Notices of the Royal Astronomical Society} \textbf{271}, 513 (1994).

\bibitem{Matarrese98} S. Matarrese, S. Mollerach, and M. Bruni: Relativistic second--order perturbations of the Einstein--de Sitter Universe. {\sl Phys. Rev. D} \textbf{58}, 
043504 (1998).

\bibitem{Mukhanov92} V.F. Mukhanov, H.A. Feldman, and R.H. Brandenberger: Theory of cosmological perturbations. {\sl Phys. Rep.} \textbf{215}, 203 (1992). 

\bibitem{odesols}
A.D. Polyanin and V.F. Zaitsev: Handbook of exact solutions for ordinary differential equations.
Boca Raton: CRC Press (1995).

\bibitem{rampfbuchert}
C. Rampf and T. Buchert: Lagrangian perturbations and the matter bispectrum I: fourth--order model for nonlinear clustering.
{\sl J.C.A.P.} \textbf{06}, 021  (2012).

\bibitem{rampf:zeldovich} 
C. Rampf and G. Rigopoulos: Zel'dovich approximation and general relativity.
{\sl Mon. Not. Roy. Astron. Soc.} \textbf{430}, L54 (2013).

\bibitem{rampfwiegand}
C. Rampf and A. Wiegand: Relativistic Lagrangian displacement field and tensor perturbations.
{\sl Phys. Rev. D} \textbf{90}, 123503 (2014).

\bibitem{rampfwong}
C. Rampf and Y.Y.Y. Wong: Lagrangian perturbations and the matter bispectrum II: the resummed one--loop correction to the matter bispectrum.
{\sl J.C.A.P.} \textbf{06}, 018 (2012).

\bibitem{russ:rza} 
H. Russ, M. Morita, M. Kasai and G. B{\"o}rner: Zel'dovich--type approximation for an inhomogeneous universe in general relativity: second--order solutions.
{\sl Phys. Rev. D}  \textbf{53}, 6881 (1996). 

\bibitem{salopek}
D.S. Salopek, J.M. Stewart and K.M. Croudace: The Zel'dovich Approximation and the relativistic Hamilton--Jacobi Equation,
{\sl Mon. Not. Roy. Astron. Soc.} \textbf{271}, 1005 (1994).

\bibitem{tomita0} K. Tomita: Non--linear theory of gravitational instability in the expanding Universe.
{\sl Prog. Theor. Phys.} \textbf{37}, 831 (1967).

\bibitem{tomita1} K. Tomita: Non--linear theory of gravitational instability in the expanding Universe. II.
{\sl Prog. Theor. Phys.} \textbf{45},  1747 (1971).

\bibitem{tomita2} K. Tomita: Non--linear theory of gravitational instability in the expanding Universe. III.
{\sl Prog. Theor. Phys.} \textbf{47}, 416 (1972).

\bibitem{zeldovich:fragmentation} 
Ya.B. Zel'dovich: Gravitational instability: An approximate theory for large density perturbations.
{\sl Astron. Astrophys.} \textbf{5}, 84 (1970).

\end{thebibliography}
\end{document}